\documentclass[twoside,11pt]{article}

\usepackage{amsmath,amsfonts}
\usepackage{amssymb,latexsym}
\usepackage{amsxtra}
\usepackage{upref}
\usepackage{exscale}
\usepackage[mathscr]{eucal}

\newcommand{\be}{\begin{equation}}
\newcommand{\ee}{\end{equation}}
\newcommand{\beq}{\begin{eqnarray}}
\newcommand{\eeq}{\end{eqnarray}}
\newcommand{\no}{\nonumber}
\newcommand{\bea}{\begin{array}}
\newcommand{\eea}{\end{array}}
\newcommand{\lb}{\label}

\newcommand{\mscr}{\mathscr}
\newcommand{\mfrak}{\mathfrak}
\newcommand{\ve}{\varepsilon}

\newcommand{\pp}{\partial}
\newcommand{\im}{\imath}
\newcommand{\ppr}{^{\boldsymbol{\prime}}}
\newcommand{\pppr}{^{\boldsymbol{\prime\prime}}}

\newcommand{\pdag}{^{\dagger}}
\newcommand{\wt}{\widetilde}
\newcommand{\ovv}{\overline}
\newcommand{\matG}{\mbox{$/\hspace*{-7pt}\hat{\mathscr{G}}$}}
\newcommand{\matW}{\mbox{$/\hspace*{-9pt}\hat{\mathscr{W}}$}}
\newcommand{\matB}{\mbox{$/\hspace*{-8pt}\hat{\mathscr{B}}$}}

\newcommand{\ph}{\phantom}
\newcommand{\scr}{\scriptstyle}
\newcommand{\scrscr}{\scriptscriptstyle}
\newcommand{\scz}{\scriptsize}

\numberwithin{equation}{section}
\allowdisplaybreaks[4]

\begin{document}

\begin{center}
{\large\bf Hubbard-Stratonovich transformations to self-energies with coset decomposition to} \vspace*{0.1cm}\\
{\large\bf anomalous pair condensates for the standard model of electroweak interactions} \vspace*{0.2cm}\\
{(Classical field theory with self-energy matrices of the irreducible propagator parts)} \vspace*{0.3cm}\\
{\bf Bernhard Mieck}\footnote{e-mail: "bjmeppstein@arcor.de"; freelance activity 2009; current location :
Zum Kohlwaldfeld 16, D-65817 Eppstein, Germany.} \vspace*{0.2cm}\\ {(article finished  at 25 November 2009)}
\end{center}
\begin{abstract}
The standard model of the strong and electroweak interactions is transformed from the ordinary path integral
with the Lagrangians of quarks and leptons and with the Abelian and non-Abelian gauge fields
to corresponding self-energies. We apply the precise formulation in terms of
massless Majorana Fermi fields with 'Nambu' doubling which naturally leads to the appropriate HST's
(Hubbard-Stratonovich-transformations) of the self-energies and to the subsequent coset decomposition for the
SSB (spontaneous symmetry breaking). The total coset decomposition of the Fermi fields is given by the dimension
\(N_{0}=90\) for the symmetry breaking \(\mbox{SO}(N_{0},N_{0})\,/\,\mbox{U}(N_{0})\otimes\mbox{U}(N_{0})\)
where the densities of fermions, related to the invariant subgroup \(\mbox{U}(N_{0})\), are contained
in a background functional for the remaining \(\mbox{SO}(N_{0},N_{0})\,/\,\mbox{U}(N_{0})\) coset
field degrees of freedom of total Fermi fields which are composed of quark and lepton pairs. We find that the Higgs field
partially enters into a sum with the self-energies for the gauge field strength tensors so that it may be very
difficult to observe pure, solely Higgs fields without a considerable contribution from the Abelian and
non-Abelian gauge field strength self-energies. The self-energy matrices of the standard model, 
representing the irreducible terms of propagation, allow {\it rigorous} derivations for the so-called "effective",
classical field theories as the Skyrme-like models from gradient expansions of the determinant remaining from
the bilinear, anomalous doubled, fermionic field integrations.\newline
\vspace*{0.1cm}

\vspace*{0.1cm}

\end{abstract}

\tableofcontents

\newpage

\section{Introduction} \lb{s1}

\subsection{The $\mbox{SO}(90,90)\,/\,\mbox{U}(90)\,\otimes\,\mbox{U}(90)$ self-energy matrix
for the total standard model} \lb{s11}

Although the standard model with the \(\mbox{SU}_{c}(3)\,\times\,\mbox{SU}_{L}(2)\,\times\,\mbox{U}_{Y}(1)\) gauge interactions contains numerous undetermined parameters, it gives profound insight into the spontaneous symmetry
breaking (SSB) of gauge symmetries with the Higgs mechanism \cite{Sred,Das1,Das2}. 
The latter phenomenon causes the various masses
of quarks and leptons which initially enter into the Lagrangian only as massless Majorana Fermi fields. The
presented treatment starts out from the path integral for the standard model with axial gauge constraints and
transforms the anti-commuting Fermi fields of leptons and quarks and also the gauge fields with their gauge
field strength tensors to corresponding self-energies after suitable HSTs (Hubbard-Stratonovich-transformations).
Apart from the Lagrangians of the standard model, we introduce symmetry breaking condensate 'seeds' for
fermionic coherent wavefunctions and for 
the even-numbered pairs of 'Nambu' or 'anomalous' doubled Fermi fields \cite{Nambu,Gold}.
The self-energies of the gauge field strength tensors only contribute as background fields in common with the
density parts of the total self-energy of the Fermi fields whereas the parts of the total self-energy, replacing
the anomalous doubled Fermi field constituents, are the remaining field degrees of freedom in a coset decomposition.
The various, prevailing transformations in this paper follow in analogy to the coset decomposition for the
strong interaction case \cite{BCSQCD}. However, one has to extend
the coset decomposition \(\mbox{SO}(N_{0},N_{0})\,/\,\mbox{U}(N_{0})\otimes\mbox{U}(N_{0})\) from the dimension
\(N_{0}=24\) for the QCD case with up-down (isospin) quark fields to \(N_{0}=90\) under inclusion of the
electroweak force. The dimension \(N_{0}=90\) is attained from counting the independent
components of quark and lepton Fermi fields with consideration of missing right-handed neutrinos. We finally
achieve the effective generating function (\ref{s1_1})
of coset matrices \(\hat{T}_{(\Psi)}(x_{p})=\exp\{-\hat{Y}(x_{p})\}\) for anomalous doubled
Fermi fields \(\Psi(x_{p})=\{\psi(x_{p})\,;\,\psi^{*}(x_{p})\}\) of leptons and quarks with the 'Nambu'
metric tensor \(\hat{\mathrm{S}}\) for the process of doubling fields \footnote{Since the precise derivation of (\ref{s1_1})
involves several types of gauge-field-'dressed' coset matrices, we mark the final, remaining coset matrix
\(\hat{T}_{(\Psi)}(x_{p})=\exp\{-\hat{Y}(x_{p})\}\) by the subscript '\(_{(\Psi)}\)' for the total,
anomalous doubled Fermi fields. The part \(\wt{\Delta}(\ldots)\) (first line in (\ref{s1_1})) denotes a
condensate 'seed' functional with a condensate 'seed' matrix \(\hat{J}_{\psi\psi;M_{\psi};N_{\psi}}^{AB}(x_{p})\)
for 'Nambu' pairs of fermions.}
\beq\lb{s1_1}
\lefteqn{Z\big[\hat{T}_{(\Psi)}(x_{p});\hat{\mscr{J}};J_{\psi};\hat{J}_{\psi\psi}\big] =
\int d[\hat{T}_{(\Psi)}^{-1}(x_{p})\;d\hat{T}_{(\Psi)}(x_{p})]\;\;\;
\wt{\Delta}\Big(\hat{T}_{(\Psi);M_{\psi};N_{\psi}}^{\boldsymbol{-1};AB}(x_{p}),
\hat{T}_{(\Psi);M_{\psi};N_{\psi}}^{AB}(x_{p});
\hat{J}_{\psi\psi;M_{\psi};N_{\psi}}^{AB}(x_{p})\Big) }  \\  \no &\times&
\mbox{DET}\Big[\hat{\mscr{M}}_{M_{\psi};N_{\psi}}^{AB}(x_{p},y_{q})\Big]^{\boldsymbol{1/2}} \;
\exp\bigg\{-\frac{\im}{2}\int_{C}d^{4}\!x_{p}\;d^{4}\!y_{q}\;\;
J_{\psi;M_{\psi}}^{T,A}(x_{p})\;\hat{\mathrm{S}}^{AB\ppr}\;
\hat{T}_{(\Psi);M_{\psi};M_{\psi}\ppr}^{B\ppr B\pppr}(x_{p})
\int_{C}d^{4}\!y_{q\ppr}\ppr  \;\times \\ \no  &\times&
\Big[\hat{\mscr{M}}_{M_{\psi}\ppr;N_{\psi}\pppr}^{\boldsymbol{-1};B\pppr A\pppr}
(x_{p},y_{q\ppr}\ppr) - \hat{1}_{M_{\psi}\ppr;N_{\psi}\pppr}^{B\pppr A\pppr}\;
\delta_{pq\ppr}\;\delta^{(4)}(x_{p}-y_{q\ppr}\ppr)\Big]\;\;
\Big(\langle\hat{\mscr{H}}_{\hat{\mathrm{S}}}\rangle^{\boldsymbol{-1}}\Big)_{N_{\psi}\pppr;N_{\psi}\ppr}^{A\pppr A\ppr}\!\!(
y_{q\ppr}\ppr,y_{q})\;\;\hat{T}_{(\Psi);N_{\psi}\ppr;N_{\psi}}^{A\ppr B}(y_{q})\;
J_{\psi;N_{\psi}}^{B}(y_{q})\bigg\}_{\mbox{;}} \\   \lb{s1_2}
\lefteqn{\hat{\mscr{M}}_{M_{\psi};N_{\psi}}^{AB}(x_{p},y_{q}) = \hat{1}_{M_{\psi};N_{\psi}}^{AB}\;
\delta_{pq}\;\delta^{(4)}(x_{p}-y_{q}) + } \\ \no \lefteqn{\hspace*{4.6cm}+
\Big[\big(\langle\hat{\mscr{H}}_{\hat{\mathrm{S}}}\rangle^{\boldsymbol{-1}}\big)\;\;
\delta\hat{\mscr{H}}_{\hat{\mathrm{S}}}\big(\hat{T}_{(\Psi)}^{-1},
\hat{T}_{(\Psi)}\big)+\big(\langle\hat{\mscr{H}}_{\hat{\mathrm{S}}}\rangle^{\boldsymbol{-1}}\big)\;\;
\hat{T}_{(\Psi)}^{\boldsymbol{-1}}\;
\hat{\mathrm{S}}\;\hat{\mscr{J}}\;\hat{T}_{(\Psi)}\Big]_{M_{\psi};N_{\psi}}^{AB}\!\!(x_{p},y_{q})  \;;}  \\  \lb{s1_3} &&
\langle\hat{\mscr{H}}_{\hat{\mathrm{S}}}\rangle = \hat{\mathrm{S}}\;\langle\hat{\mscr{H}}\rangle \;;  \\  \lb{s1_4}
\lefteqn{\delta\hat{\mscr{H}}_{\hat{\mathrm{S}}}\big(\hat{T}_{(\Psi)}^{-1},\hat{T}_{(\Psi)}\big) =
\Big(\hat{T}_{(\Psi)}^{\boldsymbol{-1}}\;\hat{\mathrm{S}}\;\langle\hat{\mscr{H}}\rangle\;\hat{T}_{(\Psi)}\Big)-
\Big(\hat{\mathrm{S}}\;\langle\hat{\mscr{H}}\rangle\Big) =
\Big(\exp\big\{\overrightarrow{\boldsymbol{[}\hat{Y}\,\boldsymbol{,}\,\ldots\boldsymbol{]_{-}}}\big\}
\hat{\mathrm{S}}\;\langle\hat{\mscr{H}}\rangle\Big) -
\Big(\hat{\mathrm{S}}\;\langle\hat{\mscr{H}}\rangle\Big) \;. }  
\eeq
The path integral (\ref{s1_1}) on the non-equilibrium time contour with the two branches '\(p,q=\pm\)'
consists of the effective kinetic part \(\langle\hat{\mscr{H}}_{\hat{\mathrm{S}}}\rangle\) with effective gauge terms
from a saddle point computation of a background averaging functional (denoted by \(\langle\ldots\rangle\))
with the self-energy "densities" of Fermi fields
as the invariant subgroup in the coset decomposition for the SSB. In correspondence to appendix D of \cite{BCSQCD},
one can perform a gradient expansion with the gradient term (\ref{s1_4}) up to fourth order for an effective
Lagrangian (Derrick's theorem \cite{raja}); hence, one has obtained a reliable, convenient, non-perturbative,
classical method for the dynamics within the total standard model from the spacetime evolution of the coset matrices
\(\hat{T}_{(\Psi)}(x_{p})\) (compare e.\ g.\ the 'Skyrme model' \cite{Brown1,Brown2}). It is also possible to calculate the
energy momentum tensor of (\ref{s1_1}-\ref{s1_4}) or of a gradient-expanded version from the infinitesimal spacetime variations
(with inclusion of the variations of the coset integration measure 
\(d[\hat{T}_{(\Psi)}^{-1}(x_{p})\:d\hat{T}_{(\Psi)}(x_{p})]\), \(\mbox{SO}(90,90)\,/\,\mbox{U}(90)\))
in order to couple to gravity; the corresponding energy-momentum tensor of (\ref{s1_1}-\ref{s1_4})
has then to act as a source tensor in the classical Einstein-field equations \cite{Hobson,Burg2}.
In this manner one can encompass all known interactions and their
non-perturbative dynamics into the spacetime evolution of the coset matrices \(\hat{T}_{(\Psi)}(x_{p})\) where
the locally Euclidean spacetime coordinates \(dx_{p}^{\mu}\) in the generating function (\ref{s1_1}) are related
by the inverse square root \(\hat{g}_{\mu\nu}^{-1/2}(x_{p})\;dx_{p}^{\mu}=dz_{p,\nu}\) of the coordinate
metric tensor \(\hat{g}_{\mu\nu}(x_{p})\) to the curved spacetime \(dz_{p,\mu}\) which are determined from
the Einstein field equations. 

According to chap.\ 5 ("Derivation of a nontrivial topology and 
the chiral anomaly") in \cite{BCSQCD}, one can also conclude for Hopf invariants \(\Pi_{3}(S^{2})=\mathsf{Z}\)
from the quaternion eigenvalues and matrix elements of the generator \(\hat{Y}(x_{p})\) within the coset matrix
\(\hat{T}_{(\Psi)}(x_{p})\); however, there occurs an anomaly cancellation within the total standard model
so that the total sum of Hopf invariants of the combined lepton and quark sectors should add to zero. This
vanishing sum of Hopf invariants \(\Pi_{3}(S^{2})=\mathsf{Z}\) should therefore constrain scattering and decays
of combined quark and lepton fields, as e.\ g.\ in neutron decay \(n\rightarrow p+e^{-}+\ovv{\nu}\).
The nontrivial topological configuration of fields can be rather involved for the case of the Hopf invariants;
as one considers the {\it pre-image} of a {\it point} from the \(S^{2}\) sphere within the internal space of fields 
to the compactified three dimensional coordinate space, one can attain closed, one dimensional loops or
toroidal configurations \cite{Manton}. 
In the case of the original Skyrme model, one assigns nontrivial configurations
or 'solitons' of homotopy \(\Pi_{3}(S^{3})=\mathsf{Z}\) to Baryons whereas 
our precise derivation \cite{BCSQCD} for the solely
strong interaction leads to nonzero Hopf invariants \(\Pi_{3}(S^{2})=\mathsf{Z}\) 
from the non-vanishing axial anomaly of QCD; in consequence, the transformed path integral 
in \cite{BCSQCD} with coset matrices is more closely
related to the Skyrme-Faddeev model \cite{Fad1}. It is therefore also of interest for the total standard model
to what extent nontrivial topological configurations (as the restrictive, vanishing sum of Hopf invariants)
can determine prevailing field combinations as baryons or mesons with the leptonic sector of electrons
and neutrions.

\section{Anomalous doubling of Fermi fields within the standard model} \lb{s2}

\subsection{Symmetry breaking source actions and anomalous pair condensates} \lb{s21}

The applied path integral for the standard model is given by contour time integrals (\ref{s2_1}) for forward
\(\eta_{p=+}=+1\) and backward \(\eta_{p=-}=-1\) propagation which is considered by the contour
time metric (\ref{s2_2}), the contour time coordinates \(x_{p=\pm}^{\mu}\) (\ref{s2_3}) and contour time derivatives
\(\hat{\pp}_{p=\pm,\mu}\) within the actions
\beq\no
\int_{C}d^{4}\!x_{p}\;\ldots &=&\int_{L^{3}}d^{3}\!\vec{x}\bigg(\int_{-\infty}^{+\infty}dx_{+}^{0}\;\ldots+
\int_{+\infty}^{-\infty}dx_{-}^{0}\;\ldots\bigg) = \int_{L^{3}}d^{3}\!\vec{x}
\bigg(\int_{-\infty}^{+\infty}dx_{+}^{0}\;\ldots-
\int_{-\infty}^{+\infty}dx_{-}^{0}\;\ldots\bigg)  \\  \lb{s2_1} &=&
\int_{L^{3}}d^{3}\!\vec{x}\bigg(\sum_{p=\pm}\int_{-\infty}^{+\infty}dx_{p}^{0}\;\eta_{p}\;\ldots\bigg) \;; \\  \lb{s2_2}
\eta_{p} &=&\big\{\underbrace{+1}_{p=+}\;;\;\underbrace{-1}_{p=-}\big\}\;; \hspace*{0.6cm}
\eta_{q} =\big\{\underbrace{+1}_{q=+}\;;\;\underbrace{-1}_{q=-}\big\}\;;\;\;\;\;
"p",\,"q"=\pm\;;  \\ \lb{s2_3} &&\hspace*{-1.8cm}
\bea{rclrclrcl}
x_{p}^{\mu}&=&\big(x_{p}^{0}\,,\,\vec{x}\big)\;; &\hspace*{0.6cm}x_{+}^{\mu}&=&
\big(x_{+}^{0}\,,\,\vec{x}\big)\;; &\hspace*{0.6cm}
x_{-}^{\mu}&=&\big(x_{-}^{0}\,,\,\vec{x}\big)\;;   \\
\hat{\pp}_{p,\mu}&=&\bigg(\frac{\pp}{\pp x_{p}^{0}}\,,\,\frac{\pp}{\pp\vec{x}}\bigg)\;; &
\hat{\pp}_{+,\mu}&=&\bigg(\frac{\pp}{\pp x_{+}^{0}}\,,\,\frac{\pp}{\pp\vec{x}}\bigg)\;; &
\hat{\pp}_{-,\mu}&=&\bigg(\frac{\pp}{\pp x_{-}^{0}}\,,\,\frac{\pp}{\pp\vec{x}}\bigg)\;.
\eea
\eeq
The appropriate description of the standard model starts out from massless Majorana fields of fermions
which obtain their corresponding masses from the spontaneous symmetry breaking with the Higgs phenomenon.
We combine the strongly interacting quark fields \(q_{m}(x_{p})\) and lepton fields \(\wt{l}_{m}(x_{p})\)
of the three basic families \(m=1,2,3\) into one single fermionic spinor field \(\psi_{m}(x_{p})\)
with the distinguishing labels \(\psi="\wt{l}",\,"q"\) where the tilde '\(\wt{\ph{l}}\)' over \(\wt{l}_{m}(x_{p})\)
indicates the missing field degree of freedom for the right-handed neutrinos \(\nu_{m,R}(x_{p})\equiv0\).
The derivation of the nonlinear sigma model \(\mbox{SO}(90,90)\,/\,\mbox{U}(90)\,\otimes\,\mbox{U}(90)\)
within this paper follows according to the formulation of Ref. \cite{Burg1} (C.P. Burgess and G.D. Moore,
"The Standard Model : A Primer"); however, we emphasize with our derivation one additional,
important point which concerns the anomalous doubling (or also 'Nambu' doubling \cite{Nambu,Gold})
within the standard model given in terms of left-handed and right-handed Majorana fields.
Therefore, we double the total, two spin-component field \(\psi_{m}(x_{p})\) (\ref{s2_4}) of leptons \(\wt{l}_{m}(x_{p})\)
and quarks \(q_{m}(x_{p})\) by their complex conjugates \(\wt{l}_{m}^{*}(x_{p})\), \(q_{m}^{*}(x_{p})\) or in total
by \(\psi_{m}^{*}(x_{p})\) which is taken into account by the
first uppercase letters \(A,B,C=1,2\) (\ref{s2_8}) of the Latin
alphabet. Hence, one has anomalous doubled Fermi fields \(\wt{L}_{m}^{A}(x_{p})\) (\ref{s2_6}),
\(Q_{m}^{A}(x_{p})\) (\ref{s2_7}) or in total
 \(\Psi_{m}^{A}(x_{p})\) (\ref{s2_5}) with the corresponding capital letters \('\wt{L}'\), \('Q'\) or \('\Psi'\)
\beq\lb{s2_4}
\psi_{m}(x_{p}) &=&\Big(\wt{l}_{m}(x_{p})\;\boldsymbol{;}
\;q_{m}(x_{p})\Big)^{T}\;;\psi="\wt{l}"\,,\,"q" \;\;; \\ \lb{s2_5}
\Psi_{m}^{A(=1,2)}(x_{p}) &=&
\Big(\overbrace{\underbrace{\wt{l}_{m}(x_{p})}_{A=1}\,,\,
\underbrace{\wt{l}_{m}^{*}(x_{p})}_{A=2}}^{\psi="\wt{l}"}\;\boldsymbol{;}\;
\overbrace{\underbrace{q_{m}(x_{p})}_{A=1}\,,\,\underbrace{q_{m}^{*}(x_{p})}_{A=2}}^{\psi="q"} \Big)^{T} \;; \\ \lb{s2_6}
\wt{L}_{m}^{A(=1,2)}(x_{p})&=&\Big(\underbrace{\wt{l}_{m}(x_{p})}_{A=1}\,,\,
\underbrace{\wt{l}_{m}^{*}(x_{p})}_{A=2}\Big)^{T}\;; \\  \lb{s2_7}
Q_{m}^{A(=1,2)}(x_{p})&=&\Big(\underbrace{q_{m}(x_{p})}_{A=1}\,,\,
\underbrace{q_{m}^{*}(x_{p})}_{A=2}\Big)^{T}\;; \\  \lb{s2_8}
A,B,C,\,\ldots &=& 1,\,2\;\;\;.
\eeq
According to the formulation of Ref. \cite{Burg1}, we point out the additional fact of the
standard model that it is entirely specified in terms of 'Nambu' or anomalous doubled Fermi
fields \(\Psi_{m}^{A}(x_{p})\), \(\wt{L}_{m}^{A}(x_{p})\), \(Q_{m}^{A}(x_{p})\) \cite{Nambu,Gold}.
In consequence this 'Nambu' doubling naturally leads to a coset decomposition for anomalous
doubled pairs of Fermi fields with a removal of the self-energy densities as background fields.

We define in relation (\ref{s2_9}) the total gauge group
\(\mbox{SU}_{c}(3)\,\otimes\,\mbox{SU}_{L}(2)\,\otimes\,\mbox{U}_{Y}(1)\)
(corresponding to Ref. \cite{Burg1}) with the first Greek letters \(\alpha,\beta,\gamma=1,\ldots,8\) for the eight gluon
fields \(G_{\mu}^{\alpha}(x_{p})\), with the first lowercase Latin letters \(a,b,c=1,2,3\) for the \(\mbox{SU}_{L}(2)\)
gauge boson fields \(W_{\mu}^{a}(x_{p})\) and with the Greek letters \(\kappa,\lambda,\mu,\nu,\rho\) of spacetime
indices for the weak hypercharge gauge boson field \(B_{\mu}(x_{p})\). Furthermore, the indices \(r,s=1,2,3\) and
\(f,g=1,2\) denote the matrix elements \((\hat{\lambda}^{\alpha})_{rs}\) of (gluon) Gell-Mann matrices and the
Pauli-iso-spin matrices \((\hat{\tau}^{a})_{fg}\) for the electroweak doublets. However, we depart from the formulation
of Ref. \cite{Burg1}
concerning the \(4\times4\) Dirac gamma matrices \((\hat{\gamma}^{\mu})_{IJ}\), (\(I,J,K=1,\ldots,4\))
and separate these into 'Nambu' doubled parts \(A,B=1,2\) consisting of \(2\times2\) Pauli spin matrices
\((\hat{\vec{\sigma}})_{i_{A}j_{B}}\) (\(i_{A},j_{B}=\uparrow,\downarrow\)) for the massless Majorana Fermi fields.
Consequently, the \(2\times2\) Pauli spin matrices within the \(4\times4\) Dirac gamma matrices transform the upper
'\(\uparrow\)' and lower '\(\downarrow\)' spin components of the Majorana fields. The particular list of definitions
is described in (\ref{s2_10}) to (\ref{s2_16})
where we especially hint to the split of the \(4\times4\) Dirac gamma matrices
\((\hat{\gamma}^{\mu})_{IJ}\), (\(I,J,K=1,\ldots,4\)) into the \(2\times2\) Pauli spin matrices for the
anomalous doubling of Fermi fields within the off-diagonal blocks '\(A\neq B\)' of
\((\hat{\gamma}^{\mu})_{i_{A}j_{B}}^{AB}\)
\beq \lb{s2_9} && \hspace*{-1.6cm}
\bea{ccccc}
\mbox{SU}_{c}(3) &\times & \mbox{SU}_{L}(2) & \times & \mbox{U}_{Y}(1)  \\
\downarrow && \downarrow && \downarrow \\
8\;G_{\mu}^{\alpha}(x_{p}) && 3\;W_{\mu}^{a}(x_{p}) && B_{\mu}(x_{p}) \\
\alpha,\,\beta,\,\gamma=1,\ldots,8 && a,\,b,\,c=1,2,3 && \kappa,\lambda,\mu,\nu,\rho=0,1,2,3   \\
\downarrow && \downarrow && \downarrow \\
(\hat{\lambda}^{\alpha})_{rs} && (\hat{\tau}^{a})_{fg} &&
(\hat{\gamma}^{\mu})_{IJ}=(\hat{\gamma}^{\mu})_{i_{A}j_{B}}^{AB} \\
r,\,s=1,2,3 && f,\,g=1,2 && I,\,J,\,K=1,2,3,4\;;  \\ &&&& A,\,B,\,C=1,2 \\
&&&&i_{A},\,j_{B},\,k_{C}=\uparrow,\downarrow\;.  \eea  \\  \lb{s2_10}
\hat{\lambda}^{\alpha} &:&\mbox{(gluon) Gell-Mann matrices}  \;; \\  \lb{s2_11}
\hat{\tau}^{a} &:& \mbox{Pauli (iso)-spin matrices of the weak interaction}  \;; \\  \lb{s2_12}
\hat{\eta}_{\mu\nu}  &:=& \mbox{diag}\big(-1\,,\,+1\,,\,+1\,,\,+1\big);\;\ve^{0123}=+1;\;
\mbox{(mostly '+' convention, cf. \cite{Burg1})}  ; \\  \lb{s2_13}
\big(\hat{\gamma}^{\mu}\big)_{IJ} &:& \hat{\gamma}^{0}=
\left(\bea{cc} 0 & -\hat{\im} \\ -\hat{\im} & 0 \eea\right)_{IJ}\;;\;\;\;
\hat{\beta}=\im\:\hat{\gamma}^{0}=\left(\bea{cc} 0 & \hat{1} \\ \hat{1} & 0 \eea\right)_{IJ}\;;\;\;\;
\hat{\vec{\gamma}}=\left(\bea{cc} 0 & -\im\,\hat{\vec{\sigma}} \\ \im\,\hat{\vec{\sigma}} & 0 \eea\right)_{IJ}\!;
{\scr(I,J=1,\ldots,4)}; \\  \no
\hat{\vec{\sigma}} &:=&\big(\hat{\sigma}_{1}\,,\,\hat{\sigma}_{2}\,,\,\hat{\sigma}_{3}\big)=
\mbox{Pauli spin-matrices}\;;  \\ \lb{s2_14}
\big(\hat{\gamma}^{\mu}\big)_{i_{A}j_{B}}^{AB} &:&
\hat{\gamma}^{0}=\left(\bea{cc} 0 & -(\hat{\im})_{i_{A}j_{B}} \\ -(\hat{\im})_{i_{A}j_{B}} & 0 \eea\right)^{AB}_{\mbox{;}}
\hat{\beta}=\im\:\big(\hat{\gamma}^{0}\big)_{i_{A}j_{B}}^{AB} \;;\;
\hat{\vec{\gamma}}=\left(\bea{cc} 0 & -\im\,(\hat{\vec{\sigma}})_{i_{A}j_{B}} \\
\im\,(\hat{\vec{\sigma}})_{i_{A}j_{B}} & 0 \eea\right)^{AB}_{\mbox{;}} \\ \no
(\hat{\vec{\sigma}})_{i_{A}j_{B}} &:=&\big((\hat{\sigma}_{1})_{i_{A}j_{B}}\,,\,
(\hat{\sigma}_{2})_{i_{A}j_{B}}\,,\,(\hat{\sigma}_{3})_{i_{A}j_{B}}\big)=
\mbox{Pauli spin-matrices}\;;\;{\scr(i_{A},j_{B}=\uparrow,\downarrow)}\;;  \\  \lb{s2_15}
\ovv{\psi}(x_{p}) &=& \psi\pdag(x_{p})\:\hat{\beta} \;; \\  \lb{s2_16}
\hat{\gamma}_{5} &:=& -\im\:\hat{\gamma}^{0}\,\hat{\gamma}^{1}\,\hat{\gamma}^{2}\,\hat{\gamma}^{3}=
\left(\bea{cc} \hat{1} & 0 \\ 0 & -\hat{1} \eea\right) ;\;  \\ \no  &&
\hat{P}_{L} = \frac{\big(\hat{1}+\hat{\gamma}_{5}\big)}{2}=
\left(\bea{cc} \hat{1} & 0 \\ 0 & 0 \eea\right) \;;\hspace*{0.3cm}
\hat{P}_{R} = \frac{\big(\hat{1}-\hat{\gamma}_{5}\big)}{2}=
\left(\bea{cc} 0 & 0 \\ 0 & \hat{1} \eea\right)_{\mbox{.}}
\eeq
In the following the detailed labeling with the various indices is defined for the lepton and quark sectors
of Fermi fields; this also allows to conclude for the dimension \(N_{0}=90\) of the coset decomposition
\(\mbox{SO}(N_{0},N_{0})\,/\,\mbox{U}(N_{0})\,\otimes\,\mbox{U}(N_{0})\)
for the total self-energy \(\mbox{SO}(N_{0},N_{0})\)
with coset matrices \(\hat{T}_{(\Psi)}(x_{p})\) \(\mbox{SO}(N_{0},N_{0})\,/\,\mbox{U}(N_{0})\) for anomalous pairs of fields and
the unitary subgroup \(\mbox{U}(N_{0})\) for self-energy densities as the vacuum or background states.
The left-handed \('H=L'\), two-component spin \('i_{A}=\uparrow,\downarrow'\) lepton fields (\ref{s2_17}-\ref{s2_20}) consist of the
three families \(m,n=1,2,3\) with the left-handed neutrino \(\wt{l}_{L}="\nu_{l}"\) and left-handed electron \(\wt{l}_{L}="e_{L}"\);
the latter distinction is also redundantly contained within the Pauli-iso-spin indices \(f,g=1,2\) for the neutrino \(f,g=1\)
and electron \(f,g=2\) where both notations will conveniently be applied in parallel. We therefore attain the dimension
\((m=1,2,3)\times(f=1,2)\times(H=L)\times(i_{A}=\uparrow,\downarrow)=3\cdot2\cdot1\cdot2=12\) within the left-handed lepton sector.
The right-handed sector of leptons (\ref{s2_19},\ref{s2_20})
lacks the right-handed neutrino field degree of freedom \(\wt{l}_{f=1,H=R}="\nu_{H=R}"\equiv0\)
with the remaining right-handed electron part (\ref{s2_20}).
As we count the contributing dimension of the right-handed lepton sector, we
are left with a total of \((m=1,2,3)\times(f=2)\times(H=R)\times(i_{A}=\uparrow,\downarrow)=3\cdot1\cdot1\cdot2=6\) components
\beq \lb{s2_17}
\wt{l}_{m,f=1,H=L,i_{A}=\uparrow,\downarrow}(x_{p}) &=& \nu_{m,H=L,i_{A}=\uparrow,\downarrow}(x_{p})\;;\;\;
(\mbox{or }\wt{l}_{L}="\nu_{L}")\;;  \\   \lb{s2_18}
\wt{l}_{m,f=2,H=L,i_{A}=\uparrow,\downarrow}(x_{p}) &=& e_{m,H=L,i_{A}=\uparrow,\downarrow}(x_{p})\;;\;\;
(\mbox{or }\wt{l}_{L}="e_{L}")\;;  \\  \lb{s2_19}
\wt{l}_{m,f=1,H=R,i_{A}=\uparrow,\downarrow}(x_{p}) &=& \nu_{m,H=R,i_{A}=\uparrow,\downarrow}(x_{p})\equiv0\;;\;\;
(\mbox{or }\wt{l}_{f=1,H=R}="\nu_{H=R}"\equiv0)\;;  \\   \lb{s2_20}
\wt{l}_{m,f=2,H=R,i_{A}=\uparrow,\downarrow}(x_{p}) &=& e_{m,H=R,i_{A}=\uparrow,\downarrow}(x_{p})\;;\;\;
(\mbox{or }\wt{l}_{R}="e_{R}")  \;.
\eeq
The electroweak doublet structure of quarks \(q="u"\), \(q="d"\) (\ref{s2_21},\ref{s2_22})
has equal numbers of right- and left-handed parts
within the three families \(m=1,2,3\) where the notation of the \(2\times2\) iso-spin matrices with \(f,g=1,2\)
equivalently refers to the \(u(p)\)- and \(d(own)\)-quark components. In comparison to the lepton sector,
one has also to include the indices \(r,s=1,2,3\) of the Gell-Mann matrices
for the gluons so that one acquires a total of
\((m=1,2,3)\times(f=1,2)\times(r=1,2,3)\times(H=L,R)\times(i_{A}=\uparrow,\downarrow)=3\cdot2\cdot3\cdot2\cdot2=72\)
components within the quark sector
\beq  \lb{s2_21}
q_{m,f=1,r=1,2,3,H=L,R,i_{A}=\uparrow,\downarrow}(x_{p}) &=& u_{m,r=1,2,3,H=L,R,i_{A}=\uparrow,\downarrow}(x_{p})\;;\;\;
(\mbox{or }q="u") \;; \\  \lb{s2_22}
q_{m,f=2,r=1,2,3,H=L,R,i_{A}=\uparrow,\downarrow}(x_{p}) &=& d_{m,r=1,2,3,H=L,R,i_{A}=\uparrow,\downarrow}(x_{p})\;;\;\;
(\mbox{or }q="d") \;.
\eeq
As we add the \(18\) components of the lepton sector to the \(72\) components of the quark sector to \(N_{0}=90\)
and as we consider the 'Nambu' metric tensor \(\hat{\mathrm{S}}^{AB}\) (\ref{s2_23})
of the anomalous doubling, one finally achieves
the total self-energy matrix \(\mbox{SO}(N_{0},N_{0})\) for the standard model to be decomposed into the coset
part \(\mbox{SO}(N_{0},N_{0})\,/\,\mbox{U}(N_{0})\) of anomalous pairs and the unitary sub-group part \(\mbox{U}(N_{0})\)
of self-energy densities. Note that the 'Nambu' metric tensor \(\hat{\mathrm{S}}^{AB}\) (\ref{s2_23})
has off-diagonal entries compared
to previous investigations \cite{pop1,pop2,BCSQCD}. This is caused by the change of the anomalous doubled density
of fields from \(\psi^{*}(x_{p})\cdot\psi(x_{p})=\frac{1}{2}\,(\Psi^{\dag,A}(x_{p})\,\hat{\mathrm{S}}^{A=B}\,\Psi^{B}(x_{p})\,)\)
with \(\hat{\mathrm{S}}^{A=B}=\mbox{diag}\{\hat{1}\,;\,-\hat{1}\}\) of Refs. \cite{pop1,pop2,BCSQCD} to the considered case
with \(\psi^{*}(x_{p})\cdot\psi(x_{p})=\frac{1}{2}\,(\Psi^{T,A}(x_{p})\,\hat{\mathrm{S}}^{A\neq B}\,\Psi^{B}(x_{p})\,)\) 
which contains the off-diagonal metric \(\hat{\mathrm{S}}^{A\neq B}\) (\ref{s2_23})
and {\it transposition} \(\Psi^{T,A}(x_{p})\) instead of the hermitian
conjugation \(\Psi^{\dag,A}(x_{p})\) as in previous investigations \cite{pop1,pop2,BCSQCD}
\be \lb{s2_23}
\hat{\mathrm{S}}^{AB}=\bigg(\bea{cc} \hat{0} & -\hat{1} \\ \hat{1} & \hat{0} \eea\bigg)^{AB} \;\;.
\ee
Apart from the Lagrangian for the dynamics of Fermi and gauge boson fields, we introduce a symmetry breaking source
action \(\mscr{A}_{S}[\hat{\mscr{J}},J_{\psi},\hat{J}_{\psi\psi}]\) (\ref{s2_24})
of the fermionic part with three source fields
\(\hat{\mscr{J}}_{M_{\psi};N_{\psi}}^{AB}(y_{q},x_{p})\), \(J_{\psi;M_{\psi}}^{A}(x_{p})\),
\(\hat{J}_{\psi\psi;M_{\psi};N_{\psi}}^{AB}(x_{p})\) where the bilinear parts have even, complex, commuting sources
\(\hat{\mscr{J}}_{M_{\psi};N_{\psi}}^{AB}(y_{q},x_{p})\), \(\hat{J}_{\psi\psi;M_{\psi};N_{\psi}}^{AB}(x_{p})\)
for anomalous doubled Fermi fields and where the linear symmetry breaking part is caused by anti-commuting, doubled
source fields \(J_{\psi;M_{\psi}}^{A}(x_{p})\) for coherent macroscopic wavefunctions
\footnote{We summarize the various indices for lepton (\ref{s2_17}-\ref{s2_20}) and quark fields (\ref{s2_21},\ref{s2_22})
by collective indices '\(M_{\psi}\)', '\(N_{\psi}\)' for brevity, cf. e.g. relations (\ref{s2_34}-\ref{s2_36}).}.
The source matrix
\(\hat{\mscr{J}}_{M_{\psi};N_{\psi}}^{AB}(y_{q},x_{p})\) is used for generating bilinear observables of Fermi
fields by differentiation
whereas the matrix field \(\hat{J}_{\psi\psi;M_{\psi};N_{\psi}}^{AB}(x_{p})\) with anti-symmetric
sub-matrices \(\hat{j}_{\psi\psi;M_{\psi};N_{\psi}}(x_{p})\), \(\hat{j}_{\psi\psi;M_{\psi};N_{\psi}}\pdag(x_{p})\)
acts as a condensate seed for anomalous paired fermionic fields after setting the matrix source field to
equivalent values \(\hat{J}_{\psi\psi;M_{\psi};N_{\psi}}^{AB}(x_{\boldsymbol{+}})=
\hat{J}_{\psi\psi;M_{\psi};N_{\psi}}^{AB}(x_{\boldsymbol{-}})\) on the two branches \(p=\pm\) of the time contour.
Similar considerations hold for the anti-commuting source field \(J_{\psi;M_{\psi}}^{A}(x_{p})\) which can be chosen
to generate an odd number of Fermi fields for observables and for condensate seeds of macroscopic wavefunctions from
equivalent values \(J_{\psi;M_{\psi}}^{A}(x_{\boldsymbol{+}})=J_{\psi;M_{\psi}}^{A}(x_{\boldsymbol{-}})\) on the
two time contour branches. In relations (\ref{s2_24}-\ref{s2_33}), we can therefore list the source action
\(\mscr{A}_{S}[\hat{\mscr{J}},J_{\psi},\hat{J}_{\psi\psi}]\) (\ref{s2_24})
with the particular property of missing right-handed
neutrinos which is separately specified for the source fields \(J_{\psi;M_{\psi}}^{A}(x_{p})\) and
\(\hat{J}_{\psi\psi;M_{\psi};N_{\psi}}^{AB}(x_{p})\) in eqs. (\ref{s2_26}-\ref{s2_33})
and is also denoted by the tilde '\(\wt{\ph{l}}\)'
over the lepton sector \(\wt{l}_{M_{\wt{l}}}(x_{p})\)
\beq  \lb{s2_24}
\lefteqn{\hspace*{-4.6cm}\mscr{A}_{S}[\hat{\mscr{J}},J_{\psi},\hat{J}_{\psi\psi}] = \frac{1}{2}\int_{C} d^{4}\!x_{p}
\bigg(J_{\psi;M_{\psi}}^{T,A}(x_{p})\;\hat{\mathrm{S}}^{AB}\;\Psi_{M_{\psi}}^{B}(x_{p})+\Psi_{M_{\psi}}^{T,A}(x_{p})\;
\hat{\mathrm{S}}^{AB}\;J_{\psi;M_{\psi}}^{B}(x_{p})\bigg)+  }    \\
\no &+&\frac{1}{2}\int_{C}d^{4}\!x_{p}\;\Psi_{M_{\psi}}^{T,A}(x_{p})\;\underbrace{\bigg(\bea{cc}
\hat{j}_{\psi\psi;M_{\psi};N_{\psi}}\pdag(x_{p}) & 0 \\ 0 & \hat{j}_{\psi\psi;M_{\psi};N_{\psi}}(x_{p})
\eea\bigg)^{AB}}_{\hat{J}_{\psi\psi;M_{\psi};N_{\psi}}^{AB}(x_{p})}\;
\Psi_{N_{\psi}}^{B}(x_{p}) + \\ \no &+&
\frac{1}{2}\int_{C}d^{4}\!x_{p}\;d^{4}\!y_{q}\;\Psi_{M_{\psi}}^{T,A}(y_{q})\;
\hat{\mscr{J}}_{M_{\psi};N_{\psi}}^{AB}(y_{q},x_{p})\;
\Psi_{N_{\psi}}^{B}(x_{p})\;; \\   \lb{s2_25}
\Psi_{M_{\psi}}^{A}(x_{p}) &=& \bigg(\wt{l}_{M_{\wt{l}}}(x_{p})\:,\:\wt{l}_{M_{\wt{l}}}^{*}(x_{p})\;\boldsymbol{;}\;
q_{M_{q}}(x_{p})\:,\:q_{M_{q}}^{*}(x_{p}) \bigg)^{T}\;; \\  \lb{s2_26}
J_{\psi;M_{\psi}}^{A}(x_{p}) &=&
\bigg(j_{\wt{l};M_{\wt{l}}}(x_{p})\:,\:j_{\wt{l};M_{\wt{l}}}^{*}(x_{p})\;\boldsymbol{;}\;
j_{q;M_{q}}(x_{p})\:,\:j_{q;M_{q}}^{*}(x_{p}) \bigg)^{T}\;; \\  \lb{s2_27}
0 &\equiv& \wt{l}_{m,f=1,H=R,i_{A}}(x_{p}) = \nu_{m,H=R,i_{A}}(x_{p}) \;;  \\  \lb{s2_28}
0 &\equiv& j_{\wt{l};m,f=1,H=R,i_{A}}=j_{\nu;m,H=R,i_{A}}(x_{p}) \;;   \\  \lb{s2_29}
j_{\wt{l}\wt{l};m,f,H_{1},i_{A};n,g,H_{2},i_{B}}(x_{p})
&=&-\;j_{\wt{l}\wt{l};m,f,H_{1},i_{A};n,g,H_{2},i_{B}}^{T}(x_{p}) \;; \\  \lb{s2_30}
j_{qq;m,f,r,H_{1},i_{A};n,g,s,H_{2},i_{B}}(x_{p})
&=&-\;j_{qq;m,f,r,H_{1},i_{A};n,g,s,H_{2},i_{B}}^{T}(x_{p}) \;; \\   \lb{s2_31}
j_{\wt{l}q;m,f,H_{1},i_{A};n,g,s,H_{2},i_{B}}(x_{p})
&=&-\;j_{\wt{l}q;m,f,H_{1},i_{A};n,g,s,H_{2},i_{B}}^{T}(x_{p}) \;; \\  \lb{s2_32}
j_{\wt{l}\wt{l};m,f=1,H_{1}=R,i_{A};n,g,H_{2},i_{B}}(x_{p})
&=& j_{\nu\wt{l};m,H_{1}=R,i_{A};n,g,H_{2},i_{B}}(x_{p}) \equiv 0 \;;   \\  \lb{s2_33}
j_{\wt{l}q;m,f=1,H_{1}=R,i_{A};n,g,s,H_{2},i_{B}}(x_{p})
&=& j_{\nu q;m,H_{1}=R,i_{A};n,g,s,H_{2},i_{B}}(x_{p}) \equiv 0 \;.
\eeq
Since one has to assign many indices for the lepton and quark sectors with its various subspaces, we have defined
collective indices \(M_{\psi}\), \(N_{\psi}\) or \(M_{\wt{l}}\), \(N_{\wt{l}}\) and
\(M_{q}\), \(N_{q}\) (\ref{s2_34}-\ref{s2_36})
with the extension that a bar over an additionally listed index, as e.g.
in \(M_{q}(\ovv{r},\ovv{i}_{A})\) (\ref{s2_36}), denotes
the omittance of these over-barred indices in the prevailing total collection of these
\beq\lb{s2_34}
M_{\psi} &:=& \bigg\{\bea{rcl}M_{\psi=\wt{l}}
&:=&\{\wt{l},m,f,H,i_{A}\}\;\;\mbox{ without right-handed neutrinos}\; \\
M_{\psi=q} &:=&\{q,m,f,r,H,i_{A}\} \eea   \\  \lb{s2_35}
N_{\psi} &:=& \bigg\{\bea{rcl}N_{\psi=\wt{l}\ppr} &:=&\{\wt{l}\ppr,n,g,H\ppr,j_{B}\}\;\;\mbox{ without right-handed neutrinos}\; \\
N_{\psi=q\ppr} &:=&\{q\ppr,n,g,s,H\ppr,j_{B}\} \eea  \\ \lb{s2_36}
M_{q}(\ovv{r},\ovv{i}_{A})
&:=& \{q,m,f,\underbrace{r}_{\mbox{\scz canceled}},H,\underbrace{i_{A}}_{\mbox{\scz canceled}}\}=
\{q,m,f,H,\}\;;\;\;\;\mbox{etc. further examples}\;.
\eeq
Similar to the source term \(\mscr{A}_{S}[\hat{\mscr{J}},J_{\psi},\hat{J}_{\psi\psi}]\) (\ref{s2_24})
of Fermi fields, we also include
a source action \(\mscr{A}_{sg}[\hat{\mfrak{j}}_{\alpha}^{(\hat{G})},
\hat{\mfrak{j}}_{a}^{(\hat{W})},\hat{\mfrak{j}}^{(\hat{B})}]\) (\ref{s2_37})
for the gauge field strength tensors with anti-symmetric, even- and real-valued source matrices
\(\hat{\mfrak{j}}^{(\hat{B})\mu\nu}(x_{p})\), \(\hat{\mfrak{j}}_{a}^{(\hat{W})\mu\nu}(x_{p})\),
\(\hat{\mfrak{j}}_{\alpha}^{(\hat{G})\mu\nu}(x_{p})\) (\ref{s2_39});
however, we omit an anomalous kind of doubling as in the case of the Fermi
fields whereas we allow for the rather general extension of the spacetime metric tensor
\(\hat{\eta}_{\mu\lambda}\;\hat{\eta}_{\nu\rho}\) (\ref{s2_38})
by corresponding \(\theta\)-terms for possible, nontrivial \(\theta\) vacua
(see the description for possible, relevant changes by these \(\theta\)-terms in Refs. \cite{Wein2})
\beq\lb{s2_37}
\mscr{A}_{sg}[\hat{\mfrak{j}}_{\alpha}^{(\hat{G})},\hat{\mfrak{j}}_{a}^{(\hat{W})},\hat{\mfrak{j}}^{(\hat{B})}] &=&
\int_{C}d^{4}\!x_{p}\bigg(\hat{\mfrak{j}}_{\alpha}^{(\hat{G})\mu\nu}(x_{p})\;\;
\hat{\eta}_{\mu\nu,\lambda\rho}^{(g_{3},\theta_{3})}\;\;
\hat{G}_{\alpha}^{\lambda\rho}(x_{p}) + \\ \no &+&
\hat{\mfrak{j}}_{a}^{(\hat{W})\mu\nu}(x_{p})\;\;\hat{\eta}_{\mu\nu,\lambda\rho}^{(g_{2},\theta_{2})}\;\;
\hat{W}_{a}^{\lambda\rho}(x_{p}) +
\hat{\mfrak{j}}^{(\hat{B})\mu\nu}(x_{p})\;\;\hat{\eta}_{\mu\nu,\lambda\rho}^{(g_{1},\theta_{1})}\;\;
\hat{B}^{\lambda\rho}(x_{p}) \bigg) \;;  \\ \lb{s2_38}
\hat{\eta}_{\mu\nu,\lambda\rho}^{(g_{1},\theta_{1})} &=& \hat{\eta}_{\mu\lambda}\;\hat{\eta}_{\nu\rho} +
\frac{g_{1}^{2}\:\theta_{1}}{16\:\pi^{2}}\;\hat{\ve}_{\mu\nu\lambda\rho} \;;\hspace*{0.7cm}
\hat{\eta}_{\mu\nu,\lambda\rho}^{(g_{2},\theta_{2})} = \hat{\eta}_{\mu\lambda}\;\hat{\eta}_{\nu\rho} +
\frac{g_{2}^{2}\:\theta_{2}}{16\:\pi^{2}}\;\hat{\ve}_{\mu\nu\lambda\rho} \;; \\  \no
\hat{\eta}_{\mu\nu,\lambda\rho}^{(g_{3},\theta_{3})} &=& \hat{\eta}_{\mu\lambda}\;\hat{\eta}_{\nu\rho} +
\frac{g_{3}^{2}\:\theta_{3}}{16\:\pi^{2}}\;\hat{\ve}_{\mu\nu\lambda\rho} \;; \\  \lb{s2_39}
\hat{\mfrak{j}}^{(\hat{B})\mu\nu}(x_{p}) &=& -\hat{\mfrak{j}}^{(\hat{B})\nu\mu}(x_{p}) \;; \hspace*{1.9cm}
\hat{\mfrak{j}}_{a}^{(\hat{W})\mu\nu}(x_{p}) = -\hat{\mfrak{j}}_{a}^{(\hat{W})\nu\mu}(x_{p})  \;; \\ \no
\hat{\mfrak{j}}_{\alpha}^{(\hat{G})\mu\nu}(x_{p}) &=& -\hat{\mfrak{j}}_{\alpha}^{(\hat{G})\nu\mu}(x_{p})  \;.
\eeq
Finally, we can state the general structure of the generating function
\(Z[\hat{\mscr{J}},J_{\psi},\hat{J}_{\psi\psi};
\hat{\mfrak{j}}^{(\hat{G})},\hat{\mfrak{j}}^{(\hat{W})},\hat{\mfrak{j}}^{(\hat{B})}]\) (\ref{s2_43})
for the standard model with the source actions
\(\mscr{A}_{S}[\hat{\mscr{J}},J_{\psi},\hat{J}_{\psi\psi}]\) (\ref{s2_24}),
\(\mscr{A}_{sg}[\hat{\mfrak{j}}_{\alpha}^{(\hat{G})},\hat{\mfrak{j}}_{a}^{(\hat{W})},
\hat{\mfrak{j}}^{(\hat{B})}]\) (\ref{s2_37}), with
the integration over the anti-commuting fields of leptons and quarks (\ref{s2_6},\ref{s2_7})
and the integration over the gauge fields
\(G_{\alpha}^{\mu}(x_{p})\), \(W^{\mu}_{a}(x_{p})\), \(B^{\mu}(x_{p})\)
whose chosen axial gauge conditions (\ref{s2_40}-\ref{s2_42}) are specified
by the auxiliary, real fields \(s_{\alpha}^{{\scrscr(G)}}(x_{p})\),
\(s_{a}^{{\scrscr(W)}}(x_{p})\), \(s^{{\scrscr(B)}}(x_{p})\) within the
standard Fourier integral representation of corresponding delta functions containing constant vectors
\(n_{{\scrscr(G)}\mu}\), \(n_{{\scrscr(W)}\mu}\) and \(n_{{\scrscr(B)}\mu}\)
\beq\lb{s2_40}
\delta\big(n_{{\scrscr(B)}\mu}\;B^{\mu}(x_{p})\,\big) &=&\int d[s^{{\scrscr(B)}}(x_{p})]\;
\exp\Big\{\im\int_{C}d^{4}\!x_{p}\;s^{{\scrscr(B)}}(x_{p})\;n_{{\scrscr(B)}\mu}\;B^{\mu}(x_{p})\Big\} \;; \\  \lb{s2_41}
\prod_{a=1}^{3}\delta\big(n_{{\scrscr(W)}\mu}\;W_{a}^{\mu}(x_{p})\,\big) &=&\int d[s_{a}^{{\scrscr(W)}}(x_{p})]\;
\exp\Big\{\im\int_{C}d^{4}\!x_{p}\;s_{a}^{{\scrscr(W)}}(x_{p})\;n_{{\scrscr(W)}}^{\mu}\;
W_{\mu}^{a}(x_{p})\Big\} \;; \\  \lb{s2_42}
\prod_{\alpha=1}^{8}\delta\big(n_{{\scrscr(G)}\mu}\;G_{\alpha}^{\mu}(x_{p})\,\big) &=&\int
d[s_{\alpha}^{{\scrscr(G)}}(x_{p})]\;
\exp\Big\{\im\int_{C}d^{4}\!x_{p}\;s_{\alpha}^{{\scrscr(G)}}(x_{p})\;
n_{{\scrscr(G)}}^{\mu}\;G_{\mu}^{\alpha}(x_{p})\Big\} \;;   \\  \lb{s2_43}
\lefteqn{\hspace*{-3.6cm}Z[\hat{\mscr{J}},J_{\psi},\hat{J}_{\psi\psi};
\hat{\mfrak{j}}^{(\hat{G})},\hat{\mfrak{j}}^{(\hat{W})},\hat{\mfrak{j}}^{(\hat{B})}] =
\int d[\wt{l}_{M_{\wt{l}}}(x_{p}),\wt{l}_{M_{\wt{l}}}^{*}(x_{p})]\;
d[q_{{\scrscr M_{q}}}(x_{p}),q_{{\scrscr M_{q}}}^{*}(x_{p})]  \;\times   }  \\ \no &\times&
\int d[\phi\pdag(x_{p}),\phi(x_{p})]\;\exp\bigg\{\im\int_{C}d^{4}\!x_{p}\;
\Big(\mscr{L}_{ff}(x_{p})+\mscr{L}_{H}(x_{p})\Big)-\im\;
\mscr{A}_{S}[\hat{\mscr{J}},J_{\psi},\hat{J}_{\psi\psi}]\bigg\} \\ \no &\times&
\int d[G_{\alpha}^{\mu}(x_{p}),s_{\alpha}^{{\scrscr(G)}}(x_{p})] \;
\int d[W_{a}^{\mu}(x_{p}),s_{a}^{{\scrscr(W)}}(x_{p})] \;\int d[B^{\mu}(x_{p}),s^{{\scrscr(B)}}(x_{p})]
\; \\ \no &\times&
\exp\bigg\{\im\int_{C}d^{4}\!x_{p}\;\Big(\mscr{L}_{gj}(x_{p})+\mscr{L}_{gg}(x_{p})\Big)-\im\;
\mscr{A}_{sg}[\hat{\mfrak{j}}_{\alpha}^{(\hat{G})},\hat{\mfrak{j}}_{a}^{(\hat{W})},
\hat{\mfrak{j}}^{(\hat{B})}]\bigg\} \;.
\eeq

\subsection{The Lagrangian of the standard model in terms of currents and auxiliary Higgs fields} \lb{s22}

It remains to describe the detailed form of the kinetic, fermionic part \(\mscr{L}_{ff}(x_{p})\) and Higgs part
\(\mscr{L}_{H}(x_{p})\) within the total Lagrangian \(\mscr{L}_{tot}(x_{p})\) where the gauge boson - current
coupling \(\mscr{L}_{gj}(x_{p})\) determines the interaction of fermions and where one also has the additional,
non-Abelian self-interaction in \(\mscr{L}_{gg}(x_{p})\) among the gauge fields
\beq\lb{s2_44}
\mscr{L}_{tot}(x_{p}) &=& \mscr{L}_{ff}(x_{p}) +
\mscr{L}_{H}(x_{p}) + \mscr{L}_{gj}(x_{p}) + \mscr{L}_{gg}(x_{p})  \;.
\eeq
Although we follow the detailed description and specification of Lagrangian parts according to \cite{Burg1},
we again outline the various parts of the standard model Lagrangian \(\mscr{L}_{tot}(x_{p})\), in order to
emphasize the natural appearance of an anomalous doubling of the Fermi fields, due to the proper formulation
with massless Majorana Fermi fields and symmetry breaking terms \cite{Nambu,Gold}. Therefore, one is guided by the formulation,
corresponding to Ref. \cite{Burg1}, to the coset decomposition into 'Nambu' doubled pairs within the self-energy and
remaining sub-group parts of the self-energy densities as background or vacuum state. We define from
\cite{Burg1} left-, right-handed, vectorial and axial gamma matrices
\((\hat{\gamma}_{H=L,R}^{\mu})_{i_{A}j_{B}}^{AB}\) (\ref{s2_45},\ref{s2_46}),
\((\hat{\gamma}_{5,H=L,R}^{\mu})_{i_{A}j_{B}}^{AB}\) (\ref{s2_47},\ref{s2_48}) which
separate into a symmetric (anti-symmetric) block structure of Pauli spin matrices in the off-diagonal blocks
\(A\neq B\) of vectorial \((\hat{\gamma}_{H=L,R}^{\mu})_{i_{A}j_{B}}^{AB}\)
(axial \((\hat{\gamma}_{5,H=L,R}^{\mu})_{i_{A}j_{B}}^{AB}\)) gamma matrices, respectively. The vectorial
gamma matrices \((\hat{\gamma}_{H=L,R}^{\mu})_{i_{A}j_{B}}^{AB}\) enter into the
kinetic part \(\mscr{L}_{ff}(x_{p})\) (\ref{s2_49})
of fermions with an additional '\(-\im\:\hat{\mathrm{S}}^{AB}\:\ve_{p}\)' term (\ref{s2_50}) for further analytic properties and
convergence of non-equilibrium Green functions
\beq \lb{s2_45}
\big(\hat{\gamma}_{H=L}^{\mu}\big)_{i_{A}j_{B}}^{AB} &=&
\left(\bea{cc} 0 & \big(-\hat{\im}\,,\,\im\:\vec{\sigma}\big)_{i_{A}j_{B}}^{\boldsymbol{T}}  \\
\big(-\hat{\im}\,,\,\im\:\vec{\sigma}\big)_{i_{A}j_{B}}  &  0 \eea\right)^{AB}  \;;  \\  \lb{s2_46}
\big(\hat{\gamma}_{H=R}^{\mu}\big)_{i_{A}j_{B}}^{AB} &=&
\left(\bea{cc} 0 & \big(-\hat{\im}\,,\,-\im\:\vec{\sigma}\big)_{i_{A}j_{B}}^{\boldsymbol{T}}  \\
\big(-\hat{\im}\,,\,-\im\:\vec{\sigma}\big)_{i_{A}j_{B}}  &  0 \eea\right)^{AB}  \;;  \\ \lb{s2_47}
\big(\hat{\gamma}_{5,H=L}^{\mu}\big)_{i_{A}j_{B}}^{AB} &=&
\left(\bea{cc} 0 & -\big(-\hat{\im}\,,\,\im\:\vec{\sigma}\big)_{i_{A}j_{B}}^{\boldsymbol{T}}  \\
\big(-\hat{\im}\,,\,\im\:\vec{\sigma}\big)_{i_{A}j_{B}}  &  0 \eea\right)^{AB}  \;;  \\  \lb{s2_48}
\big(\hat{\gamma}_{5,H=R}^{\mu}\big)_{i_{A}j_{B}}^{AB} &=&
\left(\bea{cc} 0 & \big(-\hat{\im}\,,\,-\im\:\vec{\sigma}\big)_{i_{A}j_{B}}^{\boldsymbol{T}}  \\
-\big(-\hat{\im}\,,\,-\im\:\vec{\sigma}\big)_{i_{A}j_{B}}  &  0 \eea\right)^{AB}  \;;  \\  \lb{s2_49}
\mscr{L}_{ff}(x_{p}) &=& -\frac{1}{2} \left(\bea{c}
\wt{l}_{M_{\wt{l}}}(x_{p}) \\ \wt{l}_{M_{\wt{l}}}^{*}(x_{p}) \eea\right)^{T,A}\hspace*{-0.28cm}
\Big(\big(\hat{\gamma}_{H}^{\mu}\big)_{i_{A}j_{B}}^{AB}\;
\hat{\pp}_{p,\mu}-\im\:\hat{\mathrm{S}}^{AB}\:\ve_{p}\:\delta_{i_{A}j_{B}}\Big)\delta_{M_{\wt{l}}(\ovv{i}_{A});N_{\wt{l}}(\ovv{j}_{B})}
\left(\bea{c} \wt{l}_{N_{\wt{l}}}(x_{p}) \\ \wt{l}_{N_{\wt{l}}}^{*}(x_{p}) \eea\right)^{B} \!\!+ \\ \no &-&\frac{1}{2}
 \left(\bea{c} q_{{\scrscr M_{q}}}(x_{p}) \\ q_{{\scrscr M_{q}}}^{*}(x_{p})
\eea\right)^{T,A}\Big(\big(\hat{\gamma}_{H}^{\mu}\big)_{i_{A}j_{B}}^{AB}\;\hat{\pp}_{p,\mu}-
\im\:\hat{\mathrm{S}}^{AB}\:\ve_{p}\:\delta_{i_{A}j_{B}}\Big)\delta_{M_{q}(\ovv{i}_{A});N_{q}(\ovv{j}_{B})}
\left(\bea{c} q_{{\scrscr N_{q}}}(x_{p})  \\ q_{{\scrscr M_{q}}}^{*}(x_{p}) \eea\right)^{B}  \\
\no &=& -\frac{1}{2}\Psi_{M_{\psi}}^{T,A}(x_{p})\;
\Big(\big(\hat{\gamma}_{H}^{\mu}\big)^{AB}_{i_{A}j_{B}}\;\hat{\pp}_{p,\mu}-\im\:
\hat{\mathrm{S}}^{AB}\:\ve_{p}\:\delta_{i_{A}j_{B}}\Big)\delta_{M_{\psi}(\ovv{i}_{A});N_{\psi}(\ovv{j}_{B})}\;
\Psi_{N_{\psi}}^{B}(x_{p})\;\;\;; \\  \lb{s2_50}
\ve_{p} &=& \eta_{p}\;\ve_{+}\;;\;\;\;\;\ve_{+}>0\;;\;\;\;\ve_{+}\rightarrow0_{+}\;\;\;.
\eeq
The Higgs part \(\mscr{L}_{H}(x_{p})\) (\ref{s2_51}), whose non-zero vacuum value of
\(\phi^{a}(x_{p})=(\phi_{1}(x_{p})\,,\,\phi_{2}(x_{p})\,)^{T}\) causes mass terms of fermions and couplings
among quarks according to the Kobayashi-Maskawa matrix \cite{Sred}, is also given in correspondence to \cite{Burg1};
however, we reformulate this part by introducing an anti-symmetric matrix 
\(\hat{\mfrak{F}}_{\psi\psi;M_{\psi};N_{\psi}}^{AB}\) (\ref{s2_52})
for the 'Nambu' doubled Fermi fields \(\Psi_{M_{\psi}}^{T,A}(x_{p})\;
\hat{\mfrak{F}}_{\psi\psi;M_{\psi};N_{\psi}}^{AB}(x_{p})\;\Psi_{N_{\psi}}^{B}(x_{p})\), similar to the kinetic
part which is also in total anti-symmetric, due to the anti-symmetric derivative operator \(\hat{\pp}_{p,\mu}\)
and symmetric, vectorial gamma matrices (\ref{s2_45},\ref{s2_46})
\beq \lb{s2_51}
\mscr{L}_{H}(x_{p}) &=& -\big(\hat{\pp}_{p,\mu}\phi(x_{p})\,\big)\pdag\big(\hat{\pp}_{p}^{\mu}\phi(x_{p})\,\big)+
\mu^{2}\:\big(\phi\pdag(x_{p})\:\phi(x_{p})\,\big)-\lambda\,\Big(\phi\pdag(x_{p})\:\phi(x_{p})\,\Big)^{2}
-\lambda\Big(\frac{\mu^{2}}{2\,\lambda}\Big)^{2} +  \\ \no &-&
\phi_{1}(x_{p})\Big(\nu_{m,L}\pdag(x_{p})\;\hat{f}_{mn}\;e_{n,R}(x_{p})+
u_{m,L}\pdag(x_{p})\;\hat{h}_{mn}\;d_{n,R}(x_{p})
-u_{m,R}\pdag(x_{p})\;\hat{g}_{mn}\;d_{n,L}(x_{p})\Big) +  \\ \no &-&
\phi_{1}^{*}(x_{p})\Big(e_{m,R}\pdag(x_{p})\;\hat{f}_{mn}\;\nu_{n,L}(x_{p})+
d_{m,R}\pdag(x_{p})\;\hat{h}_{mn}\;u_{n,L}(x_{p})
-d_{m,L}\pdag(x_{p})\;\hat{g}_{mn}\;u_{n,R}(x_{p})\Big) +  \\ \no &-&
\phi_{2}(x_{p})\Big(e_{m,L}\pdag(x_{p})\;\hat{f}_{mn}\;e_{n,R}(x_{p})+
d_{m,L}\pdag(x_{p})\;\hat{h}_{mn}\;d_{n,R}(x_{p})
+u_{m,R}\pdag(x_{p})\;\hat{g}_{mn}\;u_{n,L}(x_{p})\Big) +  \\ \no &-&
\phi_{2}^{*}(x_{p})\Big(e_{m,R}\pdag(x_{p})\;\hat{f}_{mn}\;e_{n,L}(x_{p})+
d_{m,R}\pdag(x_{p})\;\hat{h}_{mn}\;d_{n,L}(x_{p})
+u_{m,L}\pdag(x_{p})\;\hat{g}_{mn}\;u_{n,R}(x_{p})\Big)_{\mbox{;}}   \\  \lb{s2_52}
\mscr{L}_{H}(x_{p}) &=& -\big(\hat{\pp}_{p,\mu}\phi(x_{p})\,\big)\pdag\cdot\big(\hat{\pp}_{p}^{\mu}\phi(x_{p})\,\big)+
\mu^{2}\:\big(\phi\pdag(x_{p})\:\cdot\:\phi(x_{p})\,\big)-\lambda\,\Big(\phi\pdag(x_{p})\:\cdot\:\phi(x_{p})\,\Big)^{2}
-\lambda\Big(\frac{\mu^{2}}{2\,\lambda}\Big)^{2} +  \\ \no &-&
\left(\bea{c} \wt{l}_{M_{\wt{l}}}(x_{p}) \\ \wt{l}_{M_{\wt{l}}}^{*}(x_{p}) \eea\right)^{T,A}
\hat{F}_{M_{\wt{l}};N_{\wt{l}}}^{AB}\left(\phi_{1}(x_{p}),\phi_{2}(x_{p});\hat{f}_{mn}\right)
\left(\bea{c} \wt{l}_{N_{\wt{l}}}(x_{p}) \\ \wt{l}_{N_{\wt{l}}}^{*}(x_{p}) \eea\right)^{B} + \\ \no &-&
\left(\bea{c} q_{{\scrscr M_{q}}}(x_{p}) \\ q_{{\scrscr M_{q}}}^{*}(x_{p}) \eea\right)^{T,A}
\bigg(\hat{H}_{M_{q};N_{q}}^{AB}\!\Big(\phi_{1}(x_{p}),\phi_{2}(x_{p});\hat{h}_{mn}\Big)+
\hat{G}_{{\scrscr M_{q}};{\scrscr N_{q}}}^{AB}\!\Big(\phi_{1}(x_{p}),\phi_{2}(x_{p});\hat{g}_{mn}\Big)\bigg)
\left(\bea{c} q_{{\scrscr N_{q}}}(x_{p})  \\ q_{{\scrscr N_{q}}}^{*}(x_{p}) \eea\right)^{B}  \\ \no &=&
-\big(\hat{\pp}_{p,\mu}\phi(x_{p})\,\big)\pdag\cdot\big(\hat{\pp}_{p}^{\mu}\phi(x_{p})\,\big)+
\mu^{2}\:\big(\phi\pdag(x_{p})\:\cdot\:\phi(x_{p})\,\big)-\lambda\,\Big(\phi\pdag(x_{p})\:\cdot\:\phi(x_{p})\,\Big)^{2}
-\lambda\Big(\frac{\mu^{2}}{2\,\lambda}\Big)^{2} +  \\ \no &-&\frac{1}{2}
\Psi_{M_{\psi}}^{T,A}(x_{p})\;\hat{\mfrak{F}}_{\psi\psi;M_{\psi};N_{\psi}}^{AB}\!
\Big(\phi_{1}(x_{p}),\phi_{2}(x_{p})\Big)\;\Psi_{N_{\psi}}^{B}(x_{p})\;.
\eeq
In analogy, the Lagrangian \(\mscr{L}_{gj}(x_{p})\) (\ref{s2_53}) with anomalous doubled Fermi fields within the various
currents (apart from the pure Higgs currents \(j_{B}^{(\phi)\mu}(x_{p})\) and \(j_{W;a}^{(\phi)\mu}(x_{p})\))
follows the principle "(transposed 'Nambu' doubled Fermi field \(\Psi_{M_{\psi}}^{T,A}(x_{p})\)) $\times$
(anti-symmetric matrix or operator)$_{M_{\psi};N_{\psi}}^{AB}$ $\times$ (anomalous doubled Fermi field
\(\Psi_{N_{\psi}}^{B}(x_{p})\))" as in \(\mscr{L}_{ff}(x_{p})\) (\ref{s2_49}) or \(\mscr{L}_{H}(x_{p})\) (\ref{s2_52})
\beq\no
\mscr{L}_{gj}(x_{p}) &=&  -\frac{g_{1}}{2}\:B_{\mu}(x_{p})\:\Big(j_{B}^{(ax.)\mu}(x_{p})+
j_{B}^{(\phi)\mu}(x_{p})\Big) -  \frac{g_{2}}{2}\:W_{\mu}^{a}(x_{p})\:
\Big(j_{W;a}^{(vec.)\mu}(x_{p})+j_{W;a}^{(ax.)\mu}(x_{p})+j_{W;a}^{(\phi)\mu}(x_{p})\Big) +  \\  \lb{s2_53}  &-&
\frac{g_{3}}{2}\:G_{\mu}^{\alpha}(x_{p})\:
\Big(j_{G;\alpha}^{(vec.)\mu}(x_{p}) + j_{G;\alpha}^{(ax.)\mu}(x_{p})\Big)\;.
\eeq
We therefore include anti-symmetric and symmetric Pauli-isospin-matrices \(\hat{\tau}_{a}^{(-)}\),
\(\hat{\tau}_{a}^{(+)}\) (\ref{s2_54}) and the eight Gell-Mann (gluon) matrices \(\hat{\lambda}_{\alpha}^{(-)}\),
\(\hat{\lambda}_{\alpha}^{(-)}\) (\ref{s2_55})
\beq \lb{s2_54}
\big(\hat{\tau}_{a}^{(-)}\big)_{fg} &=&\Big(\frac{\hat{\tau}_{a}-\hat{\tau}_{a}^{T}}{2}\Big)_{fg} \;; \hspace*{1.0cm}
\big(\hat{\tau}_{a}^{(+)}\big)_{fg} =\Big(\frac{\hat{\tau}_{a}+\hat{\tau}_{a}^{T}}{2}\Big)_{fg}  \;; \\  \lb{s2_55}
\big(\hat{\lambda}_{\alpha}^{(-)}\big)_{rs} &=&
\Big(\frac{\hat{\lambda}_{\alpha}-\hat{\lambda}_{\alpha}^{T}}{2}\Big)_{rs} \;; \hspace*{1.0cm}
\big(\hat{\lambda}_{\alpha}^{(+)}\big)_{rs} =\Big(\frac{\hat{\lambda}_{\alpha}+\hat{\lambda}_{\alpha}^{T}}{2}\Big)_{rs} \;,
\eeq
which combine with the vectorial and axial gamma matrices \((\hat{\gamma}_{H}^{\mu})\) (\ref{s2_45},\ref{s2_46}),
\((\hat{\gamma}_{5,H}^{\mu})\) (\ref{s2_47},\ref{s2_48})
to completely anti-symmetric matrix couplings among the bilinear anomalous doubled Fermi fields. Relations
(\ref{s2_56}-\ref{s2_65}) encompass all the various currents of the standard model (according to Ref. \cite{Burg1})
which, however, are
entirely reformulated in terms of bilinear, anomalous doubled Fermi fields coupled to anti-symmetric matrices.
Furthermore, one has to incorporate the weak hyper-charges \(e_{\wt{l};H}^{(Y)}\), \(e_{q;H}^{(Y)}\) (\ref{s2_58})
and left- and right-handed charge values \(e_{q;H}^{(G)}\) (\ref{s2_65}) of strongly interacting quarks, in order to attain
the proper couplings for the standard model. Note that one has only left-handed current components within
\(j_{W;a}^{(vec.)\mu}(x_{p})\) (\ref{s2_60}), \(j_{W;a}^{(ax.)\mu}(x_{p})\) (\ref{s2_61}) according to the \(\mbox{SU}_{L}(2)\) gauge
group of weak interactions
\beq\lb{s2_56}
j_{B}^{(vec.)\mu}(x_{p}) &\equiv& 0\;; \\   \lb{s2_57}
j_{B}^{(ax.)\mu}(x_{p}) &=&
-\frac{1}{2}
\left(\bea{c} \wt{l}_{M_{\wt{l}}}(x_{p}) \\ \wt{l}_{M_{\wt{l}}}^{*}(x_{p}) \eea\right)^{T,A}\;\im\;
\big(\hat{\gamma}_{5,H}^{\mu}\big)_{i_{A}j_{B}}^{AB}\;
e_{\tilde{l},H}^{(Y)}\;\delta_{M_{\wt{l}}(\ovv{i}_{A});N_{\wt{l}}(j_{B})}
\left(\bea{c} \wt{l}_{N_{\wt{l}}}(x_{p}) \\ \wt{l}_{N_{\wt{l}}}^{*}(x_{p}) \eea\right)^{B} + \\ \no &-&\frac{1}{2}
\left(\bea{c} q_{{\scrscr M_{q}}}(x_{p}) \\ q_{{\scrscr M_{q}}}^{*}(x_{p}) \eea\right)^{T,A}\;\im\;
\big(\hat{\gamma}_{5,H}^{\mu}\big)_{i_{A}j_{B}}^{AB}\;
e_{q,H}^{(Y)}\;\delta_{{\scrscr M_{q}(\ovv{i}_{A})};{\scrscr N_{q}(\ovv{j}_{B})}}
\left(\bea{c} q_{{\scrscr N_{q}}}(x_{p})  \\ q_{{\scrscr N_{q}}}^{*}(x_{p}) \eea\right)   \\  \no &=&
-\frac{1}{2}\Psi_{M_{\psi}}^{T,A}(x_{p})\;\im\;\big(\hat{\gamma}_{5,H}^{\mu}\big)_{i_{A}j_{B}}^{AB}\;
e_{\psi,H}^{(Y)}\;\delta_{{\scrscr M_{\psi}(\ovv{i}_{A})};{\scrscr N_{\psi}(\ovv{j}_{B})}}\Psi_{N_{\psi}}^{B}(x_{p})\;;  \\ \lb{s2_58}
\big(e_{\tilde{l},H=L}^{(Y)}=-1&;&e_{\tilde{l},H=R}^{(Y)}=+2\big)\;;\;\;\;
\big(e_{q,H=L}^{(Y)}=+1/3\;;\;e_{q=u,H=R}^{(Y)}=-4/3\;;\;e_{q=d,H=R}^{(Y)}=+2/3\big) \;; \\  \lb{s2_59}
j_{B}^{(\phi)\mu}(x_{p}) &=& \im\Big[\phi\pdag(x_{p})\:\cdot\:\big(\hat{\pp}_{p}^{\mu}\phi(x_{p})\,\big)-
\big(\hat{\pp}_{p}^{\mu}\phi(x_{p})\,\big)\pdag\:\cdot\:\phi(x_{p})\Big] \;; \\ \lb{s2_60}
j_{W;a}^{(vec.)\mu}(x_{p}) &=& -\frac{1}{2}
\left(\bea{c} \wt{l}_{{\scrscr M_{\wt{l}}}}(x_{p}) \\ \wt{l}_{{\scrscr M_{\wt{l}}}}^{*}(x_{p}) \eea\right)^{T,A}\;\im\;
\big(\hat{\tau}_{a}^{(-)}\big)_{fg}\;\big(\hat{\gamma}_{L}^{\mu}\big)_{i_{A}j_{B}}^{AB}\;
\delta_{{\scrscr M_{\wt{l}}(\ovv{f},\ovv{i}_{A})};
{\scrscr N_{\wt{l}}(\ovv{g},\ovv{j}_{B})}}
\left(\bea{c} \wt{l}_{{\scrscr N_{\wt{l}}}}(x_{p}) \\ \wt{l}_{{\scrscr N_{\wt{l}}}}^{*}(x_{p}) \eea\right)^{B} +  \\ \no &-&\frac{1}{2}
\left(\bea{c} q_{{\scrscr M_{q}}}(x_{p}) \\ q_{{\scrscr M_{q}}}^{*}(x_{p}) \eea\right)^{T,A}\;\im\;
\big(\hat{\tau}_{a}^{(-)}\big)_{fg}\;\big(\hat{\gamma}_{L}^{\mu}\big)_{i_{A}j_{B}}^{AB}\;
\delta{{\scrscr M_{q}(\ovv{f},\ovv{i}_{A})};{\scrscr N_{q}(\ovv{g},\ovv{j}_{B})}}
\left(\bea{c} q_{{\scrscr N_{q}}}(x_{p}) \\ q_{{\scrscr N_{q}}}^{*}(x_{p}) \eea\right)^{B} = \\ \no &=&-\frac{1}{2}
\Psi_{{\scrscr M_{\psi}}}^{T,A}(x_{p})\;\im\;
\big(\hat{\tau}_{a}^{(-)}\big)_{fg}\;\big(\hat{\gamma}_{L}^{\mu}\big)_{i_{A}j_{B}}^{AB}\;
\delta_{{\scrscr M_{\psi}(\ovv{f},\ovv{i}_{A})};{\scrscr N_{\psi}(\ovv{g},\ovv{j}_{B})}}\;
\Psi_{{\scrscr N_{\psi}}}^{B}(x_{p})\;;   \\  \lb{s2_61}
j_{W;a}^{(ax.)\mu}(x_{p}) &=&  -\frac{1}{2}
\left(\bea{c} \wt{l}_{{\scrscr M_{\wt{l}}}}(x_{p}) \\ \wt{l}_{{\scrscr M_{\wt{l}}}}^{*}(x_{p}) \eea\right)^{T,A}\;\im\;
\big(\hat{\tau}_{a}^{(+)}\big)_{fg}\;\big(\hat{\gamma}_{5,L}^{\mu}\big)_{i_{A}j_{B}}^{AB}\;
\delta_{{\scrscr M_{\wt{l}}(\ovv{f},\ovv{i}_{A})};
{\scrscr N_{\wt{l}}(\ovv{g},\ovv{j}_{B})}}
\left(\bea{c} \wt{l}_{{\scrscr N_{\wt{l}}}}(x_{p}) \\ \wt{l}_{{\scrscr N_{\wt{l}}}}^{*}(x_{p}) \eea\right)^{B} +  \\ \no &-&\frac{1}{2}
\left(\bea{c} q_{{\scrscr M_{q}}}(x_{p}) \\ q_{{\scrscr M_{q}}}^{*}(x_{p}) \eea\right)^{T,A}\;\im\;
\big(\hat{\tau}_{a}^{(+)}\big)_{fg}\;\big(\hat{\gamma}_{5,L}^{\mu}\big)_{i_{A}j_{B}}^{AB}\;
\delta{{\scrscr M_{q}(\ovv{f},\ovv{i}_{A})};{\scrscr N_{q}(\ovv{g},\ovv{j}_{B})}}
\left(\bea{c} q_{{\scrscr N_{q}}}(x_{p}) \\ q_{{\scrscr N_{q}}}^{*}(x_{p}) \eea\right)^{B} = \\ \no &=&-\frac{1}{2}
\Psi_{{\scrscr M_{\psi}}}^{T,A}(x_{p})\;\im\;
\big(\hat{\tau}_{a}^{(+)}\big)_{fg}\;\big(\hat{\gamma}_{5,L}^{\mu}\big)_{i_{A}j_{B}}^{AB}\;
\delta_{{\scrscr M_{\psi}(\ovv{f},\ovv{i}_{A})};{\scrscr N_{\psi}(\ovv{g},\ovv{j}_{B})}}\;
\Psi_{{\scrscr N_{\psi}}}^{B}(x_{p})\;;   \\  \lb{s2_62}
j_{W;a}^{(\phi)\mu}(x_{p}) &=& \im\Big[\phi\pdag(x_{p})\,\hat{\tau}_{a}\,\big(\hat{\pp}_{p}^{\mu}\phi(x_{p})\,\big)-
\big(\hat{\pp}_{p}^{\mu}\phi(x_{p})\,\big)\pdag\,\hat{\tau}_{a}\,\phi(x_{p})\Big]\;;  \\   \lb{s2_63}
j_{G;\alpha}^{(vec.)\mu}(x_{p}) &=& -\frac{1}{2}
\left(\bea{c} q_{{\scrscr M_{q}}}(x_{p}) \\ q_{{\scrscr M_{q}}}^{*}(x_{p}) \eea\right)^{T,A}\;\im\;
\big(\hat{\lambda}_{\alpha}^{(-)}\big)_{rs}\;\big(\hat{\gamma}_{H}^{\mu}\big)_{i_{A}j_{B}}^{AB}\;
\delta_{M_{q}(\ovv{r},\ovv{i}_{A});N_{q}(\ovv{s},\ovv{j}_{B})}
\left(\bea{c} q_{{\scrscr N_{q}}}(x_{p}) \\ q_{{\scrscr N_{q}}}^{*}(x_{p})  \eea\right)^{B}   \\ \no &=&
-\frac{1}{2}
Q_{M_{q}}^{T,A}(x_{p})\;\im\;\big(\hat{\lambda}_{\alpha}^{(-)}\big)_{rs}\;\big(\hat{\gamma}_{H}^{\mu}\big)_{i_{A}j_{B}}^{AB}\;
\delta_{M_{q}(\ovv{r},\ovv{i}_{A});N_{q}(\ovv{s},\ovv{j}_{B})}\;Q_{N_{q}}^{B}(x_{p})  \;; \\  \lb{s2_64}
j_{G;\alpha}^{(ax.)\mu}(x_{p}) &=&  -\frac{1}{2}
\left(\bea{c} q_{{\scrscr M_{q}}}(x_{p}) \\ q_{{\scrscr M_{q}}}^{*}(x_{p}) \eea\right)^{T,A}\;\im\;
\big(\hat{\lambda}_{\alpha}^{(+)}\big)_{rs}\;\big(\hat{\gamma}_{5,H}^{\mu}\big)_{i_{A}j_{B}}^{AB}\;
\delta_{M_{q}(\ovv{r},\ovv{i}_{A});N_{q}(\ovv{s},\ovv{j}_{B})}
\left(\bea{c} q_{{\scrscr N_{q}}}(x_{p}) \\ q_{{\scrscr N_{q}}}^{*}(x_{p})  \eea\right)^{B}   \\ \no &=&
-\frac{1}{2}
Q_{M_{q}}^{T,A}(x_{p})\;\im\;\big(\hat{\lambda}_{\alpha}^{(+)}\big)_{rs}\;\big(\hat{\gamma}_{5,H}^{\mu}\big)_{i_{A}j_{B}}^{AB}\;
\delta_{M_{q}(\ovv{r},\ovv{i}_{A});N_{q}(\ovv{s},\ovv{j}_{B})}\;Q_{N_{q}}^{B}(x_{p})  \;;
\\ \lb{s2_65} && \big(e_{q;H=L}^{(G)}=+1\;,\;e_{q;H=R}^{(G)}=-1\big)\;.
\eeq
The Lagrangian \(\mscr{L}_{gg}(x_{p})\) with quadratic terms of the gauge field strength tensors
\(\hat{G}_{\mu\nu}^{\alpha}(x_{p})\) (\ref{s2_67}), \(\hat{W}_{\mu\nu}^{a}(x_{p})\) (\ref{s2_68}),
\(\hat{B}_{\mu\nu}(x_{p})\) (\ref{s2_69})
is given in relation (\ref{s2_66}) with the generalized metric tensors
\(\hat{\eta}_{\mu\nu,\lambda\rho}^{(g_{1},\theta_{1})}\),
\(\hat{\eta}_{\mu\nu,\lambda\rho}^{(g_{2},\theta_{2})}\),
\(\hat{\eta}_{\mu\nu,\lambda\rho}^{(g_{3},\theta_{3})}\) (\ref{s2_38}) of $\theta$-terms, the gauge couplings
to the Higgs field densities \((\phi\pdag(x_{p})\cdot\phi(x_{p})\,)\),
\((\phi\pdag(x_{p})\,\hat{\tau}_{a}\,\phi(x_{p})\,)\) and the auxiliary, real fields
\(s_{\alpha}^{(G)}(x_{p})\), \(s_{a}^{(W)}(x_{p})\), \(s^{(B)}(x_{p})\) and constant vectors
\(n_{(G)}^{\mu}\), \(n_{(W)}^{\mu}\), \(n_{(B)}^{\mu}\) for axial gauge fixing, respectively.
Furthermore, we hint to the various coupling parameters \(g_{3}\), \(g_{2}\), \(g_{1}\) of the strong,
weak and hyper \(\mbox{U}_{Y}(1)\) interactions which have already occurred in the coupling
\(\mscr{L}_{gj}(x_{p})\) (\ref{s2_53}) of the gauge boson fields \(G_{\mu}^{\alpha}(x_{p})\), \(W_{\mu}^{a}(x_{p})\),
\(B_{\mu}(x_{p})\) to their corresponding currents
\beq\lb{s2_66}
\mscr{L}_{gg}(x_{p}) &=& -\frac{1}{4}\,\hat{G}^{\alpha\mu\nu}(x_{p})\;
\hat{\eta}_{\mu\nu,\lambda\rho}^{(g_{1},\theta_{1})}\;\hat{G}^{\alpha\lambda\rho}(x_{p}) + \\ \no &-&
\frac{1}{4}\,\hat{W}^{a\mu\nu}(x_{p})\;
\hat{\eta}_{\mu\nu,\lambda\rho}^{(g_{2},\theta_{2})}
\hat{W}^{a\lambda\rho}(x_{p}) - \frac{g_{2}^{2}}{4}\,W_{\mu}^{a}(x_{p})\:W_{a}^{\mu}(x_{p})\;
\big(\phi\pdag(x_{p})\:\phi(x_{p})\,\big)  +
\\ \no &-& \frac{1}{4}\,\hat{B}^{\mu\nu}(x_{p})\;
\hat{\eta}_{\mu\nu,\lambda\rho}^{(g_{3},\theta_{3})}\;\hat{B}^{\lambda\rho}(x_{p}) -
\frac{g_{1}^{2}}{4}\,B_{\mu}(x_{p})\:B^{\mu}(x_{p})\;
\big(\phi\pdag(x_{p})\:\phi(x_{p})\,\big) +  \\ \no  &-&  \frac{g_{1}\:g_{2}}{2}\:
B^{\mu}(x_{p})\;W_{\mu}^{a}(x_{p})\;\big(\phi\pdag(x_{p})\;\hat{\tau}_{a}\;\phi(x_{p})\,\big)  +\\ \no &+&
G_{\mu}^{\alpha}(x_{p})\:s_{\alpha}^{\scrscr(G)}(x_{p})\:n^{\mu}_{\scrscr(G)}+
W_{\mu}^{a}(x_{p})\:s_{a}^{\scrscr(W)}(x_{p})\:n_{\scrscr(W)}^{\mu}+B_{\mu}(x_{p})\:
s^{\scrscr(B)}(x_{p})\:n_{\scrscr(B)}^{\mu}  \;;   \\ \lb{s2_67}
\hat{G}_{\mu\nu}^{\alpha}(x_{p}) &=&\hat{\pp}_{p,\mu}G_{\nu}^{\alpha}(x_{p})-
\hat{\pp}_{p,\nu}G_{\mu}^{\alpha}(x_{p})+g_{3}\;f_{\ph{\alpha}\beta\gamma}^{\alpha}\;
G_{\mu}^{\beta}(x_{p})\;G_{\nu}^{\gamma}(x_{p})\;;  \\  \lb{s2_68}
\hat{W}_{\mu\nu}^{a}(x_{p}) &=&\hat{\pp}_{p,\mu}W_{\nu}^{a}(x_{p})-
\hat{\pp}_{p,\nu}W_{\mu}^{a}(x_{p})+g_{2}\;\ve_{\ph{a}bc}^{a}\;
W_{\mu}^{b}(x_{p})\;W_{\nu}^{c}(x_{p})\;;  \\   \lb{s2_69}
\hat{B}_{\mu\nu}(x_{p}) &=& \hat{\pp}_{p,\mu}B_{\nu}(x_{p})-
\hat{\pp}_{p,\nu}B_{\mu}(x_{p}) \;\;\;.
\eeq

\section{Self-energies of gauge field strength tensors and quartic Fermi fields} \lb{s3}

\subsection{Gaussian transformations with self-energy matrices of the gauge fields} \lb{s31}

One might infer that proper HSTs with field strength tensors (\ref{s2_67}-\ref{s2_69})
of gauge fields cannot simplify the gauge field
interactions, due to the self-interaction of three and four vertex contributions of the non-Abelian
gauge fields \(G_{\mu}^{\alpha}(x_{p})\), \(W_{\mu}^{a}(x_{p})\) \cite{Muta}; however, we have already demonstrated in
Ref. \cite{BCSQCD} for the strong \(\mbox{SU}_{c}(3)\) interaction of gluon fields that one has never to use more
than the reproducing property of Gaussian Fourier transformations (the 'HST') in order to disentangle
the gluon interactions. One starts out from the Gaussian identities (\ref{s3_1}-\ref{s3_3}) for each gauge field strength tensor
\(\hat{G}_{\alpha}^{\mu\nu}(x_{p})\) (\ref{s2_67}), \(\hat{W}_{a}^{\mu\nu}(x_{p})\) (\ref{s2_68}),
\(\hat{B}^{\mu\nu}(x_{p})\) (\ref{s2_69})
with analogous source fields \(\hat{\mfrak{j}}_{\alpha}^{(\hat{G})\mu\nu}(x_{p})\),
\(\hat{\mfrak{j}}_{a}^{(\hat{W})\mu\nu}(x_{p})\), \(\hat{\mfrak{j}}^{(\hat{B})\mu\nu}(x_{p})\) and introduces
the complementary self-energies \(\hat{\mfrak{S}}_{\alpha\mu\nu}^{(\hat{G})}(x_{p})\),
\(\hat{\mfrak{S}}_{a\mu\nu}^{(\hat{W})}(x_{p})\), \(\hat{\mfrak{S}}_{\mu\nu}^{(\hat{B})}(x_{p})\),
as Gaussian integration variables with anti-symmetric spacetime indices.
In consequence the quadratic parts of the
field strength tensors with linear coupling to source fields (first two lines in (\ref{s3_4})) transform to the action (\ref{s3_5})
with the quadratic self-energies and remaining linear couplings between gauge field strength tensors and
corresponding self-energies (last line in (\ref{s3_4}))
\beq \lb{s3_1}
1&\equiv&\int d[\hat{\mfrak{S}}_{\alpha\mu\nu}^{(\hat{G})}(x_{p})]\;
\exp\bigg\{\frac{\im}{4}\int_{C}d^{\! 4}\!x_{p}\Big(\hat{\mfrak{S}}_{\alpha}^{(\hat{G})\mu\nu}(x_{p})-
\hat{G}_{\alpha}^{\mu\nu}(x_{p})-2\;\hat{\mfrak{j}}_{\alpha}^{(\hat{G})\mu\nu}(x_{p})\Big) \times \\ \no &\times&
\hat{\eta}_{\mu\nu,\lambda\rho}^{(g_{3},\theta_{3})}\:
\Big(\hat{\mfrak{S}}_{\alpha}^{(\hat{G})\lambda\rho}(x_{p})-\hat{G}_{\alpha}^{\lambda\rho}(x_{p})-
2\;\hat{\mfrak{j}}_{\alpha}^{(\hat{G})\lambda\rho}(x_{p})\Big)\bigg\}\; ; \\   \lb{s3_2}
1&\equiv&\int d[\hat{\mfrak{S}}_{a\mu\nu}^{(\hat{W})}(x_{p})]\;
\exp\bigg\{\frac{\im}{4}\int_{C}d^{\! 4}\!x_{p}\Big(\hat{\mfrak{S}}_{a}^{(\hat{W})\mu\nu}(x_{p})-
\hat{W}_{a}^{\mu\nu}(x_{p})-2\;\hat{\mfrak{j}}_{a}^{(\hat{W})\mu\nu}(x_{p})\Big) \times \\ \no &\times&
\hat{\eta}_{\mu\nu,\lambda\rho}^{(g_{2},\theta_{2})}\:
\Big(\hat{\mfrak{S}}_{a}^{(\hat{W})\lambda\rho}(x_{p})-\hat{W}_{a}^{\lambda\rho}(x_{p})-
2\;\hat{\mfrak{j}}_{a}^{(\hat{W})\lambda\rho}(x_{p})\Big)\bigg\}\; ;    \\    \lb{s3_3}
1&\equiv&\int d[\hat{\mfrak{S}}_{\mu\nu}^{(\hat{B})}(x_{p})]\;
\exp\bigg\{\frac{\im}{4}\int_{C}d^{\! 4}\!x_{p}\Big(\hat{\mfrak{S}}^{(\hat{B})\mu\nu}(x_{p})-
\hat{B}^{\mu\nu}(x_{p})-2\;\hat{\mfrak{j}}^{(\hat{B})\mu\nu}(x_{p})\Big) \times \\ \no &\times&
\hat{\eta}_{\mu\nu,\lambda\rho}^{(g_{1},\theta_{1})}\:
\Big(\hat{\mfrak{S}}^{(\hat{B})\lambda\rho}(x_{p})-\hat{B}^{\lambda\rho}(x_{p})-
2\;\hat{\mfrak{j}}^{(\hat{B})\lambda\rho}(x_{p})\Big)\bigg\}\; ;    \\     \lb{s3_4}
\lefteqn{\exp\bigg\{-\im\int_{C}d^{\! 4}\!x_{p}\;\bigg(\hat{G}_{\alpha}^{\mu\nu}(x_{p})\;
\frac{\hat{\eta}_{\mu\nu,\lambda\rho}^{(g_{3},\theta_{3})}}{4} \hat{G}_{\alpha}^{\lambda\rho}(x_{p})+
\hat{\mfrak{j}}_{\alpha}^{(\hat{G})\mu\nu}(x_{p})\:\hat{\eta}_{\mu\nu,\lambda\rho}^{(g_{3},\theta_{3})}\:
\hat{G}_{\alpha}^{\lambda\rho}(x_{p}) + \hat{W}_{a}^{\mu\nu}(x_{p})\;
\frac{\hat{\eta}_{\mu\nu,\lambda\rho}^{(g_{2},\theta_{2})}}{4} \hat{W}_{a}^{\lambda\rho}(x_{p}) +  }  \\  \no  &+&
\hat{\mfrak{j}}_{a}^{(\hat{W})\mu\nu}(x_{p})\:\hat{\eta}_{\mu\nu,\lambda\rho}^{(g_{2},\theta_{2})}\:
\hat{W}_{a}^{\lambda\rho}(x_{p}) + \hat{B}^{\mu\nu}(x_{p})\;
\frac{\hat{\eta}_{\mu\nu,\lambda\rho}^{(g_{1},\theta_{1})}}{4} \hat{B}^{\lambda\rho}(x_{p}) +
\hat{\mfrak{j}}^{(\hat{B})\mu\nu}(x_{p})\:\hat{\eta}_{\mu\nu,\lambda\rho}^{(g_{1},\theta_{1})}\:
\hat{B}^{\lambda\rho}(x_{p})  \bigg)
\bigg\}=    \\  \no &=& \int d[\hat{\mfrak{S}}_{\alpha\mu\nu}^{(\hat{G})}(x_{p})] \;
\int d[\hat{\mfrak{S}}_{a\mu\nu}^{(\hat{W})}(x_{p})] \;
\int d[\hat{\mfrak{S}}_{\mu\nu}^{(\hat{B})}(x_{p})] \;\times
\exp\bigg\{\im\;
\mscr{A}\left(\hat{\mfrak{S}}_{\alpha}^{(\hat{G})},\hat{\mfrak{j}}_{\alpha}^{(\hat{G})};
\hat{\mfrak{S}}_{a}^{(\hat{W})},\hat{\mfrak{j}}_{a}^{(\hat{W})};
\hat{\mfrak{S}}^{(\hat{B})},\hat{\mfrak{j}}^{(\hat{B})}\right) +  \\ \no &-&\frac{\im}{2}\int_{C}d^{4}\!x_{p}
\bigg(\hat{G}_{\alpha}^{\mu\nu}(x_{p})\;\hat{\eta}_{\mu\nu,\lambda\rho}^{(g_{3},\theta_{3})}\;
\hat{\mfrak{S}}_{\alpha}^{(\hat{G})\lambda\rho}(x_{p}) +
\hat{W}_{a}^{\mu\nu}(x_{p})\;
\hat{\eta}_{\mu\nu,\lambda\rho}^{(g_{2},\theta_{2})}\;\hat{\mfrak{S}}_{a}^{(\hat{W})\lambda\rho}(x_{p}) +
\hat{B}^{\mu\nu}(x_{p})\;\hat{\eta}_{\mu\nu,\lambda\rho}^{(g_{1},\theta_{1})}
\hat{\mfrak{S}}^{(\hat{B})\lambda\rho}(x_{p}) \bigg)\bigg\}_{\mbox{;}}         \\         \lb{s3_5}
\lefteqn{\mscr{A}\left(\hat{\mfrak{S}}_{\alpha}^{(\hat{G})},\hat{\mfrak{j}}_{\alpha}^{(\hat{G})};
\hat{\mfrak{S}}_{a}^{(\hat{W})},\hat{\mfrak{j}}_{a}^{(\hat{W})};
\hat{\mfrak{S}}^{(\hat{B})},\hat{\mfrak{j}}^{(\hat{B})}\right)
= \int_{C}d^{\! 4}\!x_{p} \bigg(\hat{\mfrak{S}}_{\alpha}^{(\hat{G})\mu\nu}(x_{p})
\frac{\hat{\eta}_{\mu\nu,\lambda\rho}^{(g_{3},\theta_{3})}}{4} \hat{\mfrak{S}}_{\alpha}^{(\hat{G})\lambda\rho}(x_{p}) +     }   \\
\no &+& \hat{\mfrak{S}}_{a}^{(\hat{W})\mu\nu}(x_{p}) \frac{\hat{\eta}_{\mu\nu,\lambda\rho}^{(g_{2},\theta_{2})}}{4}
\hat{\mfrak{S}}_{a}^{(\hat{W})\lambda\rho}(x_{p}) + \hat{\mfrak{S}}^{(\hat{B})\mu\nu}(x_{p})
\frac{\hat{\eta}_{\mu\nu,\lambda\rho}^{(g_{1},\theta_{1})}}{4} \hat{\mfrak{S}}^{(\hat{B})\lambda\rho}(x_{p}) +  \\ \no  &-&
\hat{\mfrak{j}}_{\alpha}^{(\hat{G})\mu\nu}(x_{p})\; \hat{\eta}_{\mu\nu,\lambda\rho}^{(g_{3},\theta_{3})}\;
\hat{\mfrak{S}}_{\alpha}^{(\hat{G})\lambda\rho}(x_{p}) - \hat{\mfrak{j}}_{a}^{(\hat{W})\mu\nu}(x_{p})\;
\hat{\eta}_{\mu\nu,\lambda\rho}^{(g_{2},\theta_{2})}\; \hat{\mfrak{S}}_{a}^{(\hat{W})\lambda\rho}(x_{p}) -
\hat{\mfrak{j}}^{(\hat{B})\mu\nu}(x_{p})\; \hat{\eta}_{\mu\nu,\lambda\rho}^{(g_{1},\theta_{1})}\;
\hat{\mfrak{S}}^{(\hat{B})\lambda\rho}(x_{p}) + \\ \no  &+&
\hat{\mfrak{j}}_{\alpha}^{(\hat{G})\mu\nu}(x_{p})\;\hat{\eta}_{\mu\nu,\lambda\rho}^{(g_{3},\theta_{3})}\;
\hat{\mfrak{j}}_{\alpha}^{(\hat{G})\lambda\rho}(x_{p})+
\hat{\mfrak{j}}_{a}^{(\hat{W})\mu\nu}(x_{p})\;\hat{\eta}_{\mu\nu,\lambda\rho}^{(g_{2},\theta_{2})}\;
\hat{\mfrak{j}}_{a}^{(\hat{W})\lambda\rho}(x_{p}) +
\hat{\mfrak{j}}^{(\hat{B})\mu\nu}(x_{p})\;\hat{\eta}_{\mu\nu,\lambda\rho}^{(g_{1},\theta_{1})}\;
\hat{\mfrak{j}}^{(\hat{B})\lambda\rho}(x_{p})  \bigg)  \;\;.
\eeq
It is the linear coupling \(\hat{G}_{\alpha}^{\mu\nu}(x_{p})\;\hat{\eta}_{\mu\nu,\lambda\rho}^{(g_{3},\theta_{3})}\;
\hat{\mfrak{S}}_{\alpha}^{(\hat{G})\lambda\rho}(x_{p})\),
\(\hat{W}_{a}^{\mu\nu}(x_{p})\;
\hat{\eta}_{\mu\nu,\lambda\rho}^{(g_{2},\theta_{2})}\;\hat{\mfrak{S}}_{a}^{(\hat{W})\lambda\rho}(x_{p})\),
\(\hat{B}^{\mu\nu}(x_{p})\;\hat{\eta}_{\mu\nu,\lambda\rho}^{(g_{1},\theta_{1})}\;
\hat{\mfrak{S}}^{(\hat{B})\lambda\rho}(x_{p})\) (last line in (\ref{s3_4})) with linear and quadratic gauge fields
\(G_{\alpha}^{\mu}(x_{p})\), \(G_{\alpha}^{\nu}(x_{p})\) in the field strength tensors
\(\hat{G}_{\alpha}^{\mu\nu}(x_{p})\) (\ref{s2_67}) and the linear coupling between gauge fields and currents that allows
to eliminate all gauge fields \(G_{\alpha}^{\mu}(x_{p})\), \(W_{a}^{\mu}(x_{p})\), \(B^{\mu}(x_{p})\)
by Gaussian integration at the expense of current-current interactions (quartic in the anomalous doubled
Fermi fields) and the occurrence of the self-energies
\(\hat{\mfrak{S}}_{\alpha\mu\nu}^{(\hat{G})}(x_{p})\),
\(\hat{\mfrak{S}}_{a\mu\nu}^{(\hat{W})}(x_{p})\), \(\hat{\mfrak{S}}_{\mu\nu}^{(\hat{B})}(x_{p})\).
We exemplify latter description to Gaussian integrations for the \(\mbox{SU}_{c}(3)\) case with the
gluon gauge fields \(G^{\beta\mu}(x_{p})\), \(G^{\gamma\nu}(x_{p})\) and add an anti-hermitian term
\(-\im\,\hat{\mfrak{e}}_{p}^{(\hat{G})}\) (\ref{s3_7}) for convergent properties
\beq   \lb{s3_6}
\lefteqn{\exp\bigg\{-\frac{\im}{2}\int_{C}d^{\! 4}\!x_{p}\;\hat{G}_{\alpha}^{\mu\nu}(x_{p})\;
\hat{\eta}_{\mu\nu,\lambda\rho}^{(g_{3},\theta_{3})}
\hat{\mfrak{S}}_{\alpha}^{(\hat{G})\lambda\rho}(x_{p})  \bigg\}=}   \\ \no &=&
\exp\bigg\{-\frac{\im}{2}\int_{C}d^{\! 4}\!x_{p}\;
\Big(\hat{\pp}_{p}^{\mu}G^{\alpha\nu}(x_{p})-\hat{\pp}_{p}^{\nu}G^{\alpha\mu}(x_{p})+
g_{3}\,f_{\ph{\alpha}\beta\gamma}^{\alpha}\;G^{\beta\mu}(x_{p})\;G^{\gamma\nu}(x_{p})\Big)\;
\hat{\eta}_{\mu\nu,\lambda\rho}^{(g_{3},\theta_{3})}
\hat{\mfrak{S}}_{\alpha}^{(\hat{G})\lambda\rho}(x_{p})  \bigg\}=  \\ \no &=&
\exp\bigg\{-\frac{\im}{2}\int_{C}d^{\! 4}\!x_{p}\;
G^{\beta\mu}(x_{p})\;
\Big[-\im\;\hat{\mfrak{e}}_{p}^{(\hat{G})}+g_{3}\,
f_{\ph{\alpha}\beta\gamma}^{\alpha}\;\hat{\eta}_{\mu\nu,\lambda\rho}^{(g_{3},\theta_{3})}
\hat{\mfrak{S}}_{\alpha}^{(\hat{G})\lambda\rho}(x_{p})\Big]\;
G^{\gamma\nu}(x_{p})\bigg\}\times  \\ \no &\times&
\exp\bigg\{-\frac{\im}{2}\int_{C}d^{\! 4}\!x_{p}
\Big(-G^{\alpha\nu}(x_{p})\;\hat{\eta}_{\mu\nu,\lambda\rho}^{(g_{3},\theta_{3})}
\big(\hat{\pp}_{p}^{\mu}\hat{\mfrak{S}}_{\alpha}^{(\hat{G})\lambda\rho}(x_{p})\big)+
G^{\alpha\mu}(x_{p})\;\hat{\eta}_{\mu\nu,\lambda\rho}^{(g_{3},\theta_{3})}
\big(\hat{\pp}_{p}^{\nu}\hat{\mfrak{S}}_{\alpha}^{(\hat{G})\lambda\rho}(x_{p})\big)\Big)\bigg\} =  \\ \no &=&
\exp\bigg\{-\frac{\im}{2}\int_{C}d^{\! 4}\!x_{p}\;
G^{\beta\mu}(x_{p})\;
\Big[-\im\;\hat{\mfrak{e}}_{p}^{(\hat{G})}+g_{3}\,
f_{\ph{\alpha}\beta\gamma}^{\alpha}\;\hat{\eta}_{\mu\nu,\lambda\rho}^{(g_{3},\theta_{3})}
\hat{\mfrak{S}}_{\alpha}^{(\hat{G})\lambda\rho}(x_{p})\Big]\;
G^{\gamma\nu}(x_{p})\bigg\}\times \\ \no &\times&
\exp\bigg\{-\im\int_{C}d^{\! 4}\!x_{p}\;G^{\alpha\mu}(x_{p})\;\hat{\eta}_{\mu\nu,\lambda\rho}^{(g_{3},\theta_{3})}
\big(\hat{\pp}_{p}^{\nu}\hat{\mfrak{S}}_{\alpha}^{(\hat{G})\lambda\rho}(x_{p})\big)\bigg\} \;; \\  \lb{s3_7}  &&
\Big(\hat{\mfrak{e}}_{p}^{(\hat{G})}\Big)_{\beta\mu,\gamma\nu} = \eta_{p}\;
\mfrak{e}_{+}^{(\hat{G})}\;\delta_{\beta\gamma}\;\delta_{\mu\nu}\;;
\hspace*{0.6cm}\mfrak{e}_{+}^{(\hat{G})}>0 \;.
\eeq
Similar transformations for the \(\mbox{SU}_{L}(2)\) gauge fields \(W^{b\mu}(x_{p})\), \(W^{c\nu}(x_{p})\)
and for \(\mbox{U}_{Y}(1)\) with \(B^{\mu}(x_{p})\), \(B^{\nu}(x_{p})\) with their gauge field
strength tensors result into the HST relations (\ref{s3_8}) with the self-energies
\(\hat{\mfrak{S}}_{\alpha\mu\nu}^{(\hat{G})}(x_{p})\),
\(\hat{\mfrak{S}}_{a\mu\nu}^{(\hat{W})}(x_{p})\),
\(\hat{\mfrak{S}}_{\mu\nu}^{(\hat{B})}(x_{p})\), being anti-symmetric in spacetime indices, and into the
auxiliary, real, axial gauge determining fields \(s_{\alpha}^{{\scrscr(G)}}(x_{p})\), \(s_{a}^{{\scrscr(W)}}(x_{p})\),
\(s^{{\scrscr(B)}}(x_{p})\) with constant vectors \(n_{{\scrscr(G)}\mu}\), \(n_{{\scrscr(W)}\mu}\), \(n_{{\scrscr(B)}\mu}\).
Therefore, one achieves following relation (\ref{s3_8}) for the gauge-gauge and gauge-fermion current Lagrangians
\(\mscr{L}_{gg}(x_{p})\) (\ref{s2_66}), \(\mscr{L}_{gj}(x_{p})\) (\ref{s2_53}) with the source action
\(\mscr{A}_{sg}[\hat{\mfrak{j}}_{\alpha}^{(\hat{G})},\hat{\mfrak{j}}_{a}^{(\hat{W})},\hat{\mfrak{j}}^{(\hat{B})}]\)
after integrating Gaussian terms of gauge fields \(G^{\beta\mu}(x_{p})\,\ldots\,G^{\gamma\nu}(x_{p})\),
\(W^{b\mu}(x_{p})\,\ldots\,W^{c\nu}(x_{p})\), \(B^{\mu}(x_{p})\,\ldots\,B^{\nu}(x_{p})\) with linear
couplings to the fermion currents and to the derivatives of the gauge field strength self-energies
\beq \lb{s3_8}
\lefteqn{\int d[G_{\alpha}^{\mu}(x_{p}),s_{\alpha}^{(G)}(x_{p})]\int d[W_{a}^{\mu}(x_{p}),s_{a}^{(W)}(x_{p})] \int
d[B^{\mu}(x_{p}),s^{(B)}(x_{p})]\;\;\times  } \\ \no &\times&
\exp\bigg\{\im\int_{C}d^{\!4}\!x_{p}\;\Big(\mscr{L}_{gg}(x_{p})+\mscr{L}_{gj}(x_{p})\Big)- \im\;
\mscr{A}_{sg}[\hat{\mfrak{j}}_{\alpha}^{(\hat{G})},\hat{\mfrak{j}}_{a}^{(\hat{W})},\hat{\mfrak{j}}^{(\hat{B})}]\bigg\}= \\
\no &=&  \int d[\hat{\mfrak{S}}_{\alpha\mu\nu}^{(\hat{G})}(x_{p}),s_{\alpha}^{(G)}(x_{p})] \; \int
d[\hat{\mfrak{S}}_{a\mu\nu}^{(\hat{W})}(x_{p}),s_{a}^{(W)}(x_{p})] \; \int
d[\hat{\mfrak{S}}_{\mu\nu}^{(\hat{B})}(x_{p}),s^{(B)}(x_{p})] \;\;\times  \\ \no &\times&
\exp\bigg\{\im\;\mscr{A}\left(\hat{\mfrak{S}}_{\alpha}^{(\hat{G})},\hat{\mfrak{j}}_{\alpha}^{(\hat{G})};
\hat{\mfrak{S}}_{a}^{(\hat{W})},\hat{\mfrak{j}}_{a}^{(\hat{W})};
\hat{\mfrak{S}}^{(\hat{B})},\hat{\mfrak{j}}^{(\hat{B})}\right)\bigg\}\;\times\;
\det\Big[\hat{\mfrak{M}}_{\beta\mu,\gamma\nu}^{(\hat{G})}
\big(\hat{\mfrak{S}}_{\alpha}^{(\hat{G})\lambda\rho}(x_{p})\,\big)
\Big]^{\boldsymbol{-1/2}} \,\times\,   \\ \no &\times&
\det\Big[\hat{\mfrak{M}}_{b\mu,c\nu}^{(\hat{W})}\big(\hat{\mfrak{S}}_{a}^{(\hat{W})\lambda\rho}(x_{p})
\,,\,\phi\pdag(x_{p})\,,\,\phi(x_{p})\,\big)\Big]^{\boldsymbol{-1/2}}\,\times \,
\det\Big[\hat{\mfrak{M}}_{\mu,\nu}^{(\hat{B})}\big(\hat{\mfrak{S}}_{a}^{(\hat{W})\lambda\rho}(x_{p})
\,,\,\phi\pdag(x_{p})\,,\,\phi(x_{p})\,\big)\Big]^{\boldsymbol{-1/2}}\times \\ \no &\times&
\exp\bigg\{\frac{\im}{2}\int_{C}d^{\! 4}\!x_{p}\;\bigg( \mfrak{J}^{(\hat{G})\beta\mu}(x_{p})\;\;
\hat{\mfrak{M}}_{\beta\mu,\gamma\nu}^{(\hat{G})\boldsymbol{-1}}
\big(\hat{\mfrak{S}}_{\alpha}^{(\hat{G})\lambda\rho}(x_{p})\,
\big)\;\;\mfrak{J}^{(\hat{G})\gamma\nu}(x_{p}) + \\  \no  &+& \mfrak{J}^{(\hat{W})b\mu}(x_{p}) \;
\hat{\mfrak{M}}_{b\mu,c\nu}^{(\hat{W})\boldsymbol{-1}}\big(\hat{\mfrak{S}}_{a}^{(\hat{W})\lambda\rho}(x_{p})
\,,\,\phi\pdag(x_{p})\,,\,\phi(x_{p})\,\big)\;\mfrak{J}^{(\hat{W})c\nu}(x_{p}) +
\\  \no &+& \mfrak{J}^{(\hat{B})\mu}(x_{p})\;
\hat{\mfrak{M}}_{\mu,\nu}^{(\hat{B})\boldsymbol{-1}}\big(\hat{\mfrak{S}}_{a}^{(\hat{W})\lambda\rho}(x_{p})
\,,\,\phi\pdag(x_{p})\,,\,\phi(x_{p})\,\big)\; \mfrak{J}^{(\hat{B})\nu}(x_{p}) \bigg)\bigg\}\;.
\eeq
Apart from the already given action \(\mscr{A}(\hat{\mfrak{S}}_{\alpha}^{(\hat{G})},
\hat{\mfrak{j}}_{\alpha}^{(\hat{G})};\hat{\mfrak{S}}_{a}^{(\hat{W})},\hat{\mfrak{j}}_{a}^{(\hat{W})};
\hat{\mfrak{S}}^{(\hat{B})},\hat{\mfrak{j}}^{(\hat{B})})\) in (\ref{s3_5}), it remains to outline the various
abbreviations of terms as the matrices \(\hat{\mfrak{M}}_{\beta\mu,\gamma\nu}^{(\hat{G})}\) (\ref{s3_9}),
\(\hat{\mfrak{M}}_{b\mu,c\nu}^{(\hat{W})}\) (\ref{s3_11}), \(\hat{\mfrak{M}}_{\mu,\nu}^{(\hat{B})}\) (\ref{s3_14}), containing
their corresponding gauge field strength self-energy \(\hat{\mfrak{S}}_{\alpha}^{(\hat{G})\lambda\rho}(x_{p})\),
\(\hat{\mfrak{S}}_{a}^{(\hat{W})\lambda\rho}(x_{p})\) and Higgs field combinations, in inverse square roots
of determinants \(\det[\ldots]^{-1/2}\) and as inverted matrices
\(\hat{\mfrak{M}}_{\beta\mu,\gamma\nu}^{(\hat{G})\boldsymbol{-1}}\),
\(\hat{\mfrak{M}}_{b\mu,c\nu}^{(\hat{W})\boldsymbol{-1}}\), \(\hat{\mfrak{M}}_{\mu,\nu}^{(\hat{B})\boldsymbol{-1}}\)
within generalized current-current interactions \(\mfrak{J}^{(\hat{G})\beta\mu}(x_{p})\,\ldots\,
\mfrak{J}^{(\hat{G})\gamma\nu}(x_{p})\), \(\mfrak{J}^{(\hat{W})b\mu}(x_{p})\,\ldots\,
\mfrak{J}^{(\hat{W})c\nu}(x_{p})\), \(\mfrak{J}^{(\hat{B})\mu}(x_{p})\,\ldots\,
\mfrak{J}^{(\hat{B})\nu}(x_{p})\). We therefore briefly list the definitions of the various abbreviations in
eqs. (\ref{s3_9}-\ref{s3_18}) where one has also to include convergence generating
epsilon terms (\ref{s3_16}-\ref{s3_18}) from the Gaussian integrations of the HST transformations
\beq \lb{s3_9}
\hat{\mfrak{M}}_{\beta\mu,\gamma\nu}^{(\hat{G})}\big(\hat{\mfrak{S}}_{\alpha}^{(\hat{G})\lambda\rho}(x_{p})\,\big) &=&
\Big[-\im\;\hat{\mfrak{e}}_{p}^{(\hat{G})}+g_{3}\,
f_{\ph{\alpha}\beta\gamma}^{\alpha}\;\hat{\eta}_{\mu\nu,\lambda\rho}^{(g_{3},\theta_{3})}
\hat{\mfrak{S}}_{\alpha}^{(\hat{G})\lambda\rho}(x_{p})\Big]  \;;   \\    \lb{s3_10}
\lefteqn{\hspace*{-5.5cm}\mfrak{J}_{\alpha\mu}^{(\hat{G})}(x_{p}) = \hat{\eta}_{\mu\nu,\lambda\rho}^{(g_{3},\theta_{3})}
\big(\hat{\pp}_{p}^{\nu}\hat{\mfrak{S}}_{\alpha}^{(\hat{G})\lambda\rho}(x_{p})\big)
-s_{\alpha}^{\scrscr(G)}(x_{p})\:n_{{\scrscr(G)}\mu}+
\frac{g_{3}}{2}\Big(j_{G;\alpha\mu}^{(vec.)}(x_{p})+j_{G;\alpha\mu}^{(ax.)}(x_{p})\Big) \;; } \\   \lb{s3_11}
\hat{\mfrak{M}}_{b\mu,c\nu}^{(\hat{W})}\big(\hat{\mfrak{S}}_{a}^{(\hat{W})\lambda\rho}(x_{p})
\,,\,\phi\pdag(x_{p})\,,\,\phi(x_{p})\,\big) &=& \Big[-\im\;\hat{\mfrak{e}}_{p}^{(\hat{W})}+g_{2}\,
\ve_{\ph{a}bc}^{a}\;\hat{\eta}_{\mu\nu,\lambda\rho}^{(g_{2},\theta_{2})}
\hat{\mfrak{S}}_{a}^{(\hat{W})\lambda\rho}(x_{p})+  \\ \no &+&
\frac{g_{2}^{2}}{2}\,\eta_{\mu\nu}\,\delta_{bc}\,\big(\phi\pdag(x_{p})\,\phi(x_{p})\,\big)\Big]  \;; \\ \lb{s3_12}
\mfrak{J}_{a\mu}^{(\hat{W})}(x_{p},\,B_{\mu}(x_{p})) &=&\mfrak{J}_{a\mu}^{(\hat{W})}(x_{p}) +
\frac{g_{1}\:g_{2}}{2}\:B_{\mu}(x_{p})\:\big(\phi\pdag(x_{p})\,\hat{\tau}_{a}\,\phi(x_{p})\,\big) \;; \\  \lb{s3_13}
\mfrak{J}_{a\mu}^{(\hat{W})}(x_{p})  &=&\bigg[\hat{\eta}_{\mu\nu,\lambda\rho}^{(g_{2},\theta_{2})}
\big(\hat{\pp}_{p}^{\nu}\hat{\mfrak{S}}_{a}^{(\hat{W})\lambda\rho}(x_{p})\big)
-s_{a}^{\scrscr(W)}(x_{p})\:n_{{\scrscr(W)}\mu}+ \\ \no &+&
\frac{g_{2}}{2}\Big(j_{W;a\mu}^{(vec.)}(x_{p})+j_{W;a\mu}^{(ax.)}(x_{p})+j_{W;a\mu}^{(\phi)}(x_{p})\Big)\bigg] \;;
\eeq
\beq  \lb{s3_14}
\hat{\mfrak{M}}_{\mu\nu}^{(\hat{B})}\big(\hat{\mfrak{S}}_{a}^{(\hat{W})\lambda\rho}(x_{p}),\phi\pdag(x_{p}),
\phi(x_{p})\big) &=&-\im\:\mfrak{e}_{p}^{(\hat{B})}\:\delta_{\mu\nu}+\frac{g_{1}^{2}}{2}\;\hat{\eta}_{\mu\nu}\;
\big(\phi\pdag(x_{p})\;\phi(x_{p})\big) + \\ \no \lefteqn{\hspace*{-5.5cm}-\frac{g_{1}^{2}\,g_{2}^{2}}{4}\;
\big(\phi\pdag(x_{p})\,\hat{\tau}^{b}\,\phi(x_{p})\big)\;\;\;
\hat{\mfrak{M}}_{b\mu,c\nu}^{(\hat{W})\boldsymbol{-1}}\big(\hat{\mfrak{S}}_{a}^{(\hat{W})\lambda\rho}(x_{p})
\,,\,\phi\pdag(x_{p})\,,\,\phi(x_{p})\,\big)\;\;\big(\phi\pdag(x_{p})\,\hat{\tau}^{c}\,\phi(x_{p})\,\big)\;;}
\eeq
\beq  \lb{s3_15}
\mfrak{J}_{\mu}^{(\hat{B})}(x_{p}) &=&\hat{\eta}_{\mu\nu,\lambda\rho}^{(g_{1},\theta_{1})}\;
\big(\hat{\pp}_{p}^{\nu}\hat{\mfrak{S}}^{(\hat{B})\lambda\rho}(x_{p})\,\big)-
s^{\scrscr(B)}(x_{p})\;n_{\scrscr{(B)};\mu} +
\frac{g_{1}}{2}\Big(j_{B;\mu}^{(ax.)}(x_{p})+j_{B;\mu}^{(\phi)}(x_{p})\Big) + \\ \no &-&
\frac{g_{1}\,g_{2}}{2}\;\big(\phi\pdag(x_{p})\,\hat{\tau}^{b}\,\phi(x_{p})\big)\;
\hat{\mfrak{M}}_{b\mu,c\nu}^{(\hat{W})\boldsymbol{-1}}\big(\hat{\mfrak{S}}_{a}^{(\hat{W})\lambda\rho}(x_{p})
\,,\,\phi\pdag(x_{p})\,,\,\phi(x_{p})\,\big)\;\mfrak{J}^{(\hat{W})c\nu}(x_{p})\;;   \\  \lb{s3_16}
\Big(\hat{\mfrak{e}}_{p}^{(\hat{G})}\Big)_{\beta\mu,\gamma\nu} &=& \eta_{p}\;
\mfrak{e}_{+}^{(\hat{G})}\;\delta_{\beta\gamma}\;\delta_{\mu\nu}\;;
\hspace*{0.6cm}\mfrak{e}_{+}^{(\hat{G})}>0 \;;  \\    \lb{s3_17}
\Big(\hat{\mfrak{e}}_{p}^{(\hat{W})}\Big)_{b\mu,c\nu} &=& \eta_{p}\;
\mfrak{e}_{+}^{(\hat{W})}\;\delta_{bc}\;\delta_{\mu\nu}\;;
\hspace*{0.6cm}\mfrak{e}_{+}^{(\hat{W})}>0 \;;    \\    \lb{s3_18}
\Big(\hat{\mfrak{e}}_{p}^{(\hat{B})}\Big)_{\mu,\nu} &=& \eta_{p}\;
\mfrak{e}_{+}^{(\hat{B})}\;\delta_{\mu\nu}\;;
\hspace*{0.6cm}\mfrak{e}_{+}^{(\hat{B})}>0 \;.
\eeq
Since the various transformations of this section appear to be involved, we 
give further details in appendices of subsequent articles for the various steps which lead to
(\ref{s3_8}) with the particular defined
parts of composed currents \(\mfrak{J}_{\alpha\mu}^{(\hat{G})}(x_{p})\) (\ref{s3_10}), 
\(\mfrak{J}_{a\mu}^{(\hat{W})}\!(x_{p},\,B_{\mu}(x_{p})\,)\) (\ref{s3_12}), 
\(\mfrak{J}_{a\mu}^{(\hat{W})}(x_{p})\) (\ref{s3_13}), \(\mfrak{J}_{\mu}^{(\hat{B})}(x_{p})\) (\ref{s3_15}) and
with the corresponding propagators 
\(\hat{\mfrak{M}}_{\beta\mu,\gamma\nu}^{(\hat{G})}\!(\hat{\mfrak{S}}_{\alpha}^{(\hat{G})\lambda\rho}(x_{p})\,)\)
(\ref{s3_9}), \(\hat{\mfrak{M}}_{b\mu,c\nu}^{(\hat{W})}\!(\hat{\mfrak{S}}_{a}^{(\hat{W})\lambda\rho}(x_{p})
\,,\,\phi\pdag(x_{p})\,,\,\phi(x_{p})\,)\) (\ref{s3_11}), 
\(\hat{\mfrak{M}}_{\mu\nu}^{(\hat{B})}\!(\hat{\mfrak{S}}_{a}^{(\hat{W})\lambda\rho}(x_{p}),\phi\pdag(x_{p}),
\phi(x_{p})\,)\) (\ref{s3_14}) for the strong, electroweak and hypercharge cases, respectively.

\section{HST and coset decomposition of anomalous doubled Fermi fields} \lb{s4}

\subsection{Transformation of current-current terms or quartic, bilinear Fermi fields to self-energies}\lb{s41}

As we combine eq. (\ref{s3_8}) for the HST to the self-energies of the gauge field strength tensors with the
fermion parts \(\mscr{L}_{ff}(x_{p})\) (\ref{s2_49}), \(\mscr{L}_{H}(x_{p})\) (\ref{s2_52}),
one acquires the path integral (\ref{s4_1}) with
anti-commuting integration variables \(d[\wt{L}_{M_{\wt{l}}}^{A}(x_{p})]\), \(d[Q_{M_{q}}^{A}(x_{p})]\)
\beq \lb{s4_1}
\lefteqn{Z[\hat{\mscr{J}},J_{\psi},\hat{J}_{\psi\psi};
\hat{\mfrak{j}}^{(\hat{G})},\hat{\mfrak{j}}^{(\hat{W})},\hat{\mfrak{j}}^{(\hat{B})}] = \int
d[\hat{\mfrak{S}}_{\alpha\mu\nu}^{(\hat{G})}(x_{p}),s_{\alpha}^{(G)}(x_{p})] \; \int
d[\hat{\mfrak{S}}_{a\mu\nu}^{(\hat{W})}(x_{p}),s_{a}^{(W)}(x_{p})] \; \int
d[\hat{\mfrak{S}}_{\mu\nu}^{(\hat{B})}(x_{p}),s^{(B)}(x_{p})] \; \times}
\\ \no &\times& \int d[\phi\pdag(x_{p}),\phi(x_{p})]\;
\exp\bigg\{\im\;\mscr{A}\left(\hat{\mfrak{S}}_{\alpha}^{(\hat{G})},\hat{\mfrak{j}}_{\alpha}^{(\hat{G})};
\hat{\mfrak{S}}_{a}^{(\hat{W})},\hat{\mfrak{j}}_{a}^{(\hat{W})};
\hat{\mfrak{S}}^{(\hat{B})},\hat{\mfrak{j}}^{(\hat{B})}\right)\bigg\}\;\times\;
\det\Big[\hat{\mfrak{M}}_{\beta\mu,\gamma\nu}^{(\hat{G})}\big(\hat{\mfrak{S}}_{\alpha}^{(\hat{G})\lambda\rho}(x_{p})\,\big)
\Big]^{\boldsymbol{-1/2}} \,\times\,   \\ \no &\times&
\det\Big[\hat{\mfrak{M}}_{b\mu,c\nu}^{(\hat{W})}\big(\hat{\mfrak{S}}_{a}^{(\hat{W})\lambda\rho}(x_{p})
\,,\,\phi\pdag(x_{p})\,,\,\phi(x_{p})\,\big)\Big]^{\boldsymbol{-1/2}}\,\times \,
\det\Big[\hat{\mfrak{M}}_{\mu,\nu}^{(\hat{B})}\big(\hat{\mfrak{S}}_{a}^{(\hat{W})\lambda\rho}(x_{p})
\,,\,\phi\pdag(x_{p})\,,\,\phi(x_{p})\,\big)\Big]^{\boldsymbol{-1/2}}\times  \\  \no &\times& \int
d[\wt{L}_{M_{\wt{l}}}^{A}(x_{p})]\;d[Q_{M_{q}}^{A}(x_{p})] \;\exp\Big\{\im\int_{C}d^{4}\!x_{p}\; \Big(\mscr{L}_{ff}(x_{p})+\mscr{L}_{H}(x_{p})\Big)-
\im\; \mscr{A}_{S}[\hat{\mscr{J}},J_{\psi},\hat{J}_{\psi\psi}]\Big\} \times  \\ \no &\times&
\exp\bigg\{\frac{\im}{2}\int_{C}d^{\! 4}\!x_{p}\;\bigg( \mfrak{J}^{(\hat{G})\beta\mu}(x_{p})\;\;
\hat{\mfrak{M}}_{\beta\mu,\gamma\nu}^{(\hat{G})\boldsymbol{-1}}\big(\hat{\mfrak{S}}_{\alpha}^{(\hat{G})\lambda\rho}(x_{p})\,
\big)\;\;\mfrak{J}^{(\hat{G})\gamma\nu}(x_{p}) + \\  \no  &+& \mfrak{J}^{(\hat{W})b\mu}(x_{p}) \;
\hat{\mfrak{M}}_{b\mu,c\nu}^{(\hat{W})\boldsymbol{-1}}\big(\hat{\mfrak{S}}_{a}^{(\hat{W})\lambda\rho}(x_{p})
\,,\,\phi\pdag(x_{p})\,,\,\phi(x_{p})\,\big)\;\mfrak{J}^{(\hat{W})c\nu}(x_{p}) +
\\  \no &+& \mfrak{J}^{(\hat{B})\mu}(x_{p})\;
\hat{\mfrak{M}}_{\mu,\nu}^{(\hat{B})\boldsymbol{-1}}\big(\hat{\mfrak{S}}_{a}^{(\hat{W})\lambda\rho}(x_{p})
\,,\,\phi\pdag(x_{p})\,,\,\phi(x_{p})\,\big)\; \mfrak{J}^{(\hat{B})\nu}(x_{p}) \bigg)\bigg\} \;.
\eeq
In the following we reorder the various terms according to their number of Fermi fields occurring in
the various actions within (\ref{s4_1}). At first we note the zero order terms of fermionic field degrees of freedom
in the first three lines of (\ref{s4_1}) which entirely consist of gauge field strength self-energies, Higgs field
combinations and the auxiliary, real gauge fixing integration variables. However, there are even more
zero order terms of Fermi fields hidden in the generalized currents
\(\mfrak{J}^{(\hat{G})\beta\mu}(x_{p})\) (\ref{s3_10}), \(\mfrak{J}^{(\hat{W})b\mu}(x_{p})\) (\ref{s3_13}) and
\(\mfrak{J}^{(\hat{B})\mu}(x_{p})\) (\ref{s3_15})
which we separate in relations (\ref{s4_2}-\ref{s4_8}) by introducing a tilde '\(\wt{\ph{\mfrak{J}}}\)'
over the parts of currents without bilinear fermionic field content
\beq  \lb{s4_2}
\mfrak{J}_{\alpha\mu}^{(\hat{G})}(x_{p}) &=& \frac{g_{3}}{2}\Big(j_{G;\alpha\mu}^{(vec.)}(x_{p})+
j_{G;\alpha\mu}^{(ax.)}(x_{p})\Big)+ \wt{\mfrak{J}}_{\alpha\mu}^{(\hat{G})}\!
\big(\hat{\mfrak{S}}_{\alpha}^{(\hat{G})\lambda\rho}(x_{p}),s_{\alpha}^{(G)}(x_{p})\big)\;;  \\   \lb{s4_3}\hspace*{-0.6cm}
\wt{\mfrak{J}}_{\alpha\mu}^{(\hat{G})}\!
\big(\hat{\mfrak{S}}_{\alpha}^{(\hat{G})\lambda\rho}(x_{p}),s_{\alpha}^{(G)}(x_{p})\big) &=&
\hat{\eta}_{\mu\nu,\lambda\rho}^{(g_{3},\theta_{3})}\;
\big(\hat{\pp}_{p}^{\nu}\hat{\mfrak{S}}_{\alpha}^{(\hat{G})\lambda\rho}(x_{p})\big)-
s_{\alpha}^{(G)}(x_{p})\;n_{(G);\mu}\;;  \\   \lb{s4_4}
\mfrak{J}_{a\mu}^{(\hat{W})}\!(x_{p}) &=& \frac{g_{2}}{2}\Big(j_{W;a\mu}^{(vec.)}(x_{p}) +
j_{W;a\mu}^{(ax.)}(x_{p})\Big)+
\wt{\mfrak{J}}_{a\mu}^{(\hat{W})}\!\big(\hat{\mfrak{S}}_{a}^{(\hat{W})\lambda\rho}(x_{p}),s_{a}^{(W)}(x_{p}),
\phi\pdag(x_{p}),\phi(x_{p})\big)  ; \\    \lb{s4_5}
\wt{\mfrak{J}}_{a\mu}^{(\hat{W})}\!(x_{p}) &=&
\wt{\mfrak{J}}_{a\mu}^{(\hat{W})}\!\big(\hat{\mfrak{S}}_{a}^{(\hat{W})\lambda\rho}(x_{p}),s_{a}^{(W)}(x_{p}),
\phi\pdag(x_{p}),\phi(x_{p})\big)  \\  \no &=&
\frac{g_{2}}{2}\;j_{W;a\mu}^{(\phi)}(x_{p}) + \hat{\eta}_{\mu\nu,\lambda\rho}^{(g_{2},\theta_{2})}\;
\big(\hat{\pp}_{p}^{\nu}\hat{\mfrak{S}}_{a}^{(\hat{W})\lambda\rho}(x_{p})\big)-s_{a}^{(W)}(x_{p})\;n_{(W);\mu}\;; \\  \lb{s4_6}
\mfrak{J}_{\mu}^{(\hat{B})}\!(x_{p}) &=& \frac{g_{1}}{2}\;j_{B;\mu}^{(ax.)}(x_{p}) +
\frac{g_{1}}{2}\;\hat{\mfrak{N}}_{\mu,c\nu}^{(\hat{W})}\!\big(\hat{\mfrak{S}}_{a}^{(\hat{W})\lambda\rho}(x_{p}),
\phi\pdag(x_{p}),\phi(x_{p})\big)\quad\mfrak{J}^{(\hat{W})c\nu}\!(x_{p}) +   \\  \no &+&
\wt{\mfrak{J}}_{\mu}^{(\hat{B})}\!\big(\hat{\mfrak{S}}^{(\hat{B})\lambda\rho}(x_{p}),s^{(B)}(x_{p}),\phi\pdag(x_{p}),
\phi(x_{p})\big) \;;  \\   \lb{s4_7}
\wt{\mfrak{J}}_{\mu}^{(\hat{B})}\!\big(x_{p}\big)  &=&
\wt{\mfrak{J}}_{\mu}^{(\hat{B})}\!\big(\hat{\mfrak{S}}^{(\hat{B})\lambda\rho}(x_{p}),s^{(B)}(x_{p}),\phi\pdag(x_{p}),
\phi(x_{p})\big)  \\ \no &=&
\frac{g_{1}}{2}\;j_{B;\mu}^{(\phi)}(x_{p}) + \hat{\eta}_{\mu\nu,\lambda\rho}^{(g_{1},\theta_{1})}\;
\big(\hat{\pp}_{p}^{\nu}\hat{\mfrak{S}}^{(\hat{B})\lambda\rho}(x_{p})\big)-s^{(B)}(x_{p})\;n_{(B);\mu} \;; \\  \lb{s4_8}
\hat{\mfrak{N}}_{\mu,c\nu}^{(\hat{W})}\!\big(x_{p}\big)  &=&
\hat{\mfrak{N}}_{\mu,c\nu}^{(\hat{W})}\!\big(\hat{\mfrak{S}}_{a}^{(\hat{W})\lambda\rho}(x_{p}),
\phi\pdag(x_{p}),\phi(x_{p})\big) \\  \no &=& -g_{2}\;\big(\phi\pdag(x_{p})\:\hat{\tau}^{b}\:\phi(x_{p})\big)\quad
\hat{\mfrak{M}}_{b\mu,c\nu}^{(\hat{W})-1}\!\big(\hat{\mfrak{S}}_{a}^{(\hat{W})\lambda\rho}(x_{p}),\phi\pdag(x_{p}),
\phi(x_{p})\big)\;.
\eeq
These zero order terms of Fermi fields, as the first three lines of (\ref{s4_1}) and the current-current interaction
with only 'tilded' currents (\ref{s4_3},\ref{s4_5},\ref{s4_7})  are only kept as background averaging actions for the anomalous
doubled parts of self-energies of Fermi fields. Aside from these zero order terms of Fermi fields, there also
appears a linear term of Fermi fields which is coupled to the symmetry breaking, 'Nambu' doubled,
anti-commuting source field \(J_{\psi;M_{\psi}}^{A}(x_{p})\) (\ref{s2_26}). Moreover, one has several bilinear
parts of 'Nambu' doubled Fermi fields which comprise the Lagrangians \(\mscr{L}_{ff}(x_{p})\) (\ref{s2_49}),
\(\mscr{L}_{H}(x_{p})\) (\ref{s2_52}), the couplings to the generating matrix \(\hat{\mscr{J}}_{M_{\psi};N_{\psi}}^{AB}(x_{p},y_{q})\)
for observables and the condensate 'seed' matrix \(\hat{J}_{\psi\psi;M_{\psi};N_{\psi}}^{AB}(x_{p})\) and which
also result from the generalized current-current interactions (last three lines of (\ref{s4_1})). The latter
bilinear Fermi fields follow from the product of 'tilded' currents (\ref{s4_3},\ref{s4_5},\ref{s4_7})
without any Fermi fields with the already
defined bilinear, fermionic currents \(j_{B}^{(ax.)\mu}(x_{p})\) (\ref{s2_57}), \(j_{W;a}^{(vec.)\mu}(x_{p})\) (\ref{s2_60}),
\(j_{W;a}^{(ax.)\mu}(x_{p})\) (\ref{s2_61}), \(j_{G;\alpha}^{(vec.)\mu}(x_{p})\) (\ref{s2_63}),
\(j_{G;\alpha}^{(ax.)\mu}(x_{p})\) (\ref{s2_64}).
These bilinear parts of Fermi fields could be directly removed by integration according to the relation of
anti-commuting variables \(\xi^{i}\) and an anti-symmetric matrix \(\hat{M}_{ij}\) (the symmetric part of
\(\hat{M}=+\hat{M}^{T}\) cancels in (\ref{s4_9})!)
\be\lb{s4_9}
\int d[\xi]\;\;\;\exp\Big\{-\xi^{i}\;\hat{M}_{ij}\;\xi^{j}\Big\}=\Big(\det\big[\hat{M}_{ij}\big]\Big)^{1/2}\;;\;\;\;
\hat{M}=-\hat{M}^{T}\;;\;\;\;({\scr i,j=1},\ldots,\mbox{\scz even-numbered dimension})\;,
\ee
if the path integral (\ref{s4_1}) did not contain the current-current interactions of purely bilinear, fermionic currents
or quartic Fermi fields. In order to eliminate these quartic, fermionic field combinations by integration,
one therefore has to perform additional HST transformations of quartic Fermi fields, similar to that in
Ref. \cite{BCSQCD} for the strong interaction, but in the present case with three different gauge field interactions and
corresponding self-energies. We skip the details of these transformations to purely bilinear Fermi fields and
refer to \cite{BCSQCD} as an example with the derivation in the QCD case 
(details will be listed in appendices of further articles in order to outline the various steps
of the HST to anomalous doubled self-energies of Fermi fields with 'hinge' fields as density terms). 
The coset decomposition to the self-energy matrix
\(\hat{T}_{{\scrscr(\Psi)}}(x_{p})\) (\ref{s4_10}-\ref{s4_12})
(\(\mbox{SO}(N_{0},N_{0})\,/\,\mbox{U}(N_{0})\); \(N_{0}=90\)), 
replacing the anomalous doubled Fermi field pairs, is obtained in analogy to the purely strong QCD interaction case of Ref. \cite{BCSQCD},
after removal of several gauge-field-'dressed' coset matrices
\beq \lb{s4_10}
\hat{T}_{{\scrscr(\Psi)M_{\psi};N_{\psi}}}^{AB}(x_{p}) &=& \Big(\exp\Big\{
-\hat{Y}_{{\scrscr(\Psi)M_{\psi}\ppr;N_{\psi}\ppr}}^{A\ppr B\ppr}(x_{p})\Big\}\Big)_{M_{\psi};N_{\psi}}^{AB}\;; \\ \lb{s4_11}
\hat{Y}_{{\scrscr(\Psi)M_{\psi}\ppr;N_{\psi}\ppr}}^{A\ppr\neq B\ppr}(x_{p}) &=&\bigg(\bea{cc} 0 &
\hat{X}_{{\scrscr(\Psi)M_{\psi}\ppr;N_{\psi}\ppr}}(x_{p})  \\
\hat{X}_{{\scrscr(\Psi)M_{\psi}\ppr;N_{\psi}\ppr}}\pdag(x_{p}) & 0 \eea\bigg)^{A\ppr B\ppr}_{\mbox{;}} \\  \lb{s4_12}
\hat{X}_{{\scrscr(\Psi)M_{\psi}\ppr;N_{\psi}\ppr}}(x_{p}) &=&
-\Big(\hat{X}_{{\scrscr(\Psi)M_{\psi}\ppr;N_{\psi}\ppr}}(x_{p})\Big)^{T} \;\in\mathbb{C}\;\;\;.
\eeq
One eventually succeeds into following path integral (\ref{s4_13}) of anomalous doubled parts from self-energy matrices
for Fermi fields within a coset decomposition under inclusion of the invariant integration measure
\(d[\hat{T}_{(\Psi)}^{-1}(x_{p})\;d\hat{T}_{(\Psi)}(x_{p})]\) of \(\mbox{SO}(N_{0},N_{0})\,/\,\mbox{U}(N_{0})\), (\(N_{0}=90\));
furthermore, the path integral consists of several background fields as the gauge field strength self-energies,
Higgs fields and unitary subgroup parts with self-energy "densities" which only substitute the 
"density" terms composed of fermions. This background averaging is marked by the brackets
\(\langle d[\hat{T}_{{\scrscr(\Psi)}}^{-1}(x_{p})\;d\hat{T}_{{\scrscr(\Psi)}}(x_{p})]\;\ldots\ldots\rangle\)
\beq\no
\lefteqn{Z\big[\hat{T}_{(\Psi)}(x_{p});\hat{\mscr{J}};J_{\psi};\hat{J}_{\psi\psi}\big] =
\boldsymbol{\bigg\langle} \int d[\hat{T}_{(\Psi)}^{-1}(x_{p})\;d\hat{T}_{(\Psi)}(x_{p})]\;\;
\wt{\Delta}\Big(\hat{T}_{(\Psi);M_{\psi};N_{\psi}}^{\boldsymbol{-1};AB}(x_{p})\,,\,
\hat{T}_{(\Psi);M_{\psi};N_{\psi}}^{AB}(x_{p})\,;\,
\hat{J}_{\psi\psi;M_{\psi};N_{\psi}}^{AB}(x_{p})\Big)\times }  \\  \lb{s4_13} &\times&
\mbox{DET}\Big[\hat{\mscr{M}}_{M_{\psi};N_{\psi}}^{AB}(x_{p},y_{q})\Big]^{1/2} \;
\exp\bigg\{-\frac{\im}{2}\int_{C}d^{4}\!x_{p}\;d^{4}\!y_{q}\;\;
J_{\psi;M_{\psi}}^{T,A}(x_{p})\;\hat{\mathrm{S}}^{AB\ppr}\;
\hat{T}_{(\Psi);M_{\psi};M_{\psi}\ppr}^{B\ppr B\pppr}(x_{p})
\int_{C}d^{4}\!y_{q\ppr}\ppr  \;\times \\ \no  &\times&
\Big[\hat{\mscr{M}}_{M_{\psi}\ppr;N_{\psi}\pppr}^{\boldsymbol{-1};B\pppr A\pppr}
(x_{p},y_{q\ppr}\ppr) - \hat{1}_{M_{\psi}\ppr;N_{\psi}\pppr}^{B\pppr A\pppr}\;
\delta_{pq\ppr}\;\delta^{(4)}(x_{p}-y_{q\ppr}\ppr)\Big]\;
\Big(\hat{\mscr{H}}_{\hat{\mathrm{S}}}^{\boldsymbol{-1}}\Big)_{N_{\psi}\pppr;N_{\psi}\ppr}^{A\pppr A\ppr}(
y_{q\ppr}\ppr,y_{q})\;\hat{T}_{(\Psi);N_{\psi}\ppr;N_{\psi}}^{A\ppr B}(y_{q})\;
J_{\psi;N_{\psi}}^{B}(y_{q})\bigg\}\boldsymbol{\bigg\rangle} \;; \\   \lb{s4_14}
\lefteqn{\hat{\mscr{M}}_{M_{\psi};N_{\psi}}^{AB}(x_{p},y_{q}) = \hat{1}_{M_{\psi};N_{\psi}}^{AB}\;
\delta_{pq}\;\delta^{(4)}(x_{p}-y_{q}) +}  \\ \no &&\hspace*{3.6cm}+
\Big[\big(\hat{\mscr{H}}_{\hat{\mathrm{S}}}^{\boldsymbol{-1}}\big)\;
\delta(\hat{\mscr{H}}_{\hat{\mathrm{S}}}(\hat{T}_{(\Psi)}^{-1},
\hat{T}_{(\Psi)})+\big(\hat{\mscr{H}}_{\hat{\mathrm{S}}}^{\boldsymbol{-1}}\big)\;\hat{T}_{(\Psi)}^{\boldsymbol{-1}}\;
\hat{\mathrm{S}}\;\hat{\mscr{J}}\;\hat{T}_{(\Psi)}\Big]_{M_{\psi};N_{\psi}}^{AB}\negthickspace(x_{p},y_{q})  \;;  \\  \lb{s4_15} &&
\hat{\mscr{H}}_{\hat{\mathrm{S}}} = \hat{\mathrm{S}}\;\hat{\mscr{H}} \;;  \\  \lb{s4_16}
\lefteqn{\delta\hat{\mscr{H}}_{\hat{\mathrm{S}}}(\hat{T}_{(\Psi)}^{-1},\hat{T}_{(\Psi)}) =
\Big(\hat{T}_{(\Psi)}^{\boldsymbol{-1}}\;\hat{\mathrm{S}}\;\hat{\mscr{H}}\;\hat{T}_{(\Psi)}\Big)-
\Big(\hat{\mathrm{S}}\;\hat{\mscr{H}}\Big) =
\Big(\exp\big\{\overrightarrow{\boldsymbol{[}\hat{Y}\,\boldsymbol{,}\,\ldots\boldsymbol{]_{-}}}\big\}
\hat{\mathrm{S}}\;\hat{\mscr{H}}\Big) -
\Big(\hat{\mathrm{S}}\;\hat{\mscr{H}}\Big) \;; }   \\   \lb{s4_17}
\lefteqn{\hat{\mscr{H}}_{M_{\psi};N_{\psi}}^{AB}(x_{p},y_{q}) =
\bigg[\Big(\big(\hat{\gamma}_{H}^{\nu}\big)_{i_{A}j_{B}}^{AB}\;\frac{\pp}{\pp x_{p}^{\nu}}
-\im\:\hat{\mathrm{S}}^{AB}\:\ve_{p}\;\delta_{i_{A}j_{B}}\Big)\;\delta_{M_{\psi}(\ovv{i}_{A});N_{\psi}(\ovv{j}_{B})}
+\hat{\mfrak{F}}_{\psi\psi;M_{\psi};N_{\psi}}^{AB}(x_{p}) +  }  \\ \no &+&
\bigl(\boldsymbol{\matG}(x_{p})\bigr)_{M_{q};N_{q}} + 
\bigl(\boldsymbol{\matW}(x_{p}) + \boldsymbol{\matB}(x_{p})\bigr)_{M_{\psi};N_{\psi}} +
\frac{1}{16}\;\hat{\mathrm{S}}^{AB\ppr}\bigg(\sum^{(\kappa)=0,\ldots,3}
\hat{\mfrak{s}}_{M_{\psi};N_{\psi}}^{(\Psi,\hat{B})(\kappa)B\ppr A\ppr}(x_{p}) + \\ \no &+&
\sum_{(a)=1,2,3}^{(\kappa)=0,\ldots,3}
\hat{\mfrak{s}}_{M_{\psi};N_{\psi}}^{(\Psi,\hat{W})(a;\kappa)B\ppr A\ppr}(x_{p}) +
\sum_{(\alpha)=1,\ldots,8}^{(\kappa)=0,\ldots,3}
\hat{\mfrak{s}}_{M_{q};N_{q}}^{(Q,\hat{G})(\alpha;\kappa)B\ppr A\ppr}(x_{p})\bigg)\;\hat{\mathrm{S}}^{A\ppr B} \bigg]\:
\eta_{p}\:\delta_{pq}\;\delta^{(4)}(x_{p}-y_{q})   \;;  \\  \lb{s4_18} &&
\boldsymbol{\matG}(x_{p}) = \delta_{\psi="q"}\;
\mscr{G}_{\nu}^{\gamma}(x_{p})\:\im\,
\Big[\big(\hat{\lambda}_{\gamma}^{(-)}\big)_{rs}\,\big(\hat{\gamma}_{H}^{\nu}\big)_{i_{A}j_{B}}^{AB}+
\big(\hat{\lambda}_{\gamma}^{(+)}\big)_{rs}\,
\big(\hat{\gamma}_{5,H}^{\nu}\big)_{i_{A}j_{B}}^{AB}\:e_{q;H}^{(G)}\Big]\;
\delta_{M_{q}(\ovv{r},\ovv{i}_{A});N_{q}(\ovv{s},\ovv{j}_{B})}\;; \\  \lb{s4_19}&&
\boldsymbol{\matW}(x_{p}) =  \mscr{W}_{\nu}^{c}(x_{p})\,\im\,
\Big[\big(\hat{\tau}_{c}^{(-)}\big)_{fg}\,\big(\hat{\gamma}_{L}^{\nu}\big)_{i_{A}j_{B}}^{AB}+
\big(\hat{\tau}_{c}^{(+)}\big)_{fg}\,\big(\hat{\gamma}_{5,L}^{\nu}\big)_{i_{A}j_{B}}^{AB}\Big]\,
\delta_{M_{\psi}(\ovv{f},\ovv{i}_{A});N_{\psi}(\ovv{g},\ovv{j}_{B})}\;; \\  \lb{s4_20} &&
\boldsymbol{\matB}(x_{p}) =  \mscr{B}_{\nu}(x_{p})\:\im\,
\big(\hat{\gamma}_{5,H}^{\nu}\big)_{i_{A}j_{B}}^{AB}\:\;e_{\psi;H}^{(Y)}\:\;
\delta_{M_{\psi}(\ovv{i}_{A});N_{\psi}(\ovv{j}_{B})} \;\;;  \\  \lb{s4_21} &&
\mscr{G}_{\nu}^{\gamma}(x_{p}) =\frac{g_{3}}{2}\:\;\wt{\mfrak{J}}_{\beta}^{(\hat{G})\mu}\big(
\hat{\mfrak{S}}_{\beta}^{(\hat{G})\lambda\rho}(x_{p}),s_{\beta}^{(G)}(x_{p})\big)\;\;
\hat{\mfrak{M}}_{\mu,\nu}^{(\hat{G})-1;\beta,\gamma}\big(\hat{\mfrak{S}}_{\lambda\rho}^{(\hat{G})\alpha}(x_{p})\big)\;\;;
\\  && \lb{s4_22}  \mscr{W}_{\nu}^{c}(x_{p}) = \frac{g_{2}}{2}\:\wt{\mfrak{J}}_{b}^{(\hat{W})\mu}(x_{p})\;\;
\hat{\mfrak{M}}_{\mu,\nu}^{(\hat{W})-1;b,c}(x_{p})+\frac{g_{1}\:g_{2}}{4}\:
\wt{\mfrak{J}}_{\mu}^{(\hat{B})}(x_{p})\;\;
\hat{\mfrak{M}}^{(\hat{W})-1;\mu,\kappa}(x_{p})\;\;\hat{\mfrak{N}}_{\kappa;\nu}^{(\hat{W})c}(x_{p})+ \\ \no &+&
\frac{g_{1}^{2}\:g_{2}}{8}\:\wt{\mfrak{J}}_{b\mu}^{(\hat{W})}(x_{p})\;\;
\hat{\mfrak{N}}_{\kappa}^{(\hat{W})(b\mu)}(x_{p})\;\;\hat{\mfrak{M}}^{(\hat{B})-1;\kappa\lambda}(x_{p})\;\;
\hat{\mfrak{N}}_{\lambda;\nu}^{(\hat{W})c}(x_{p})\;\;;  \\  \lb{s4_23} &&
\mscr{B}_{\nu}(x_{p}) =  \frac{g_{1}}{2}\:\wt{\mfrak{J}}^{(\hat{B})\mu}(x_{p})\;\;
\hat{\mfrak{M}}_{\mu,\nu}^{(\hat{W})-1}(x_{p})+\frac{g_{1}^{2}}{4}\:
\wt{\mfrak{J}}^{(\hat{W})b\kappa}(x_{p})\;\;\hat{\mfrak{N}}_{(b\kappa)}^{(\hat{W})\mu}(x_{p})\;\;
\hat{\mfrak{M}}_{\mu,\nu}^{(\hat{B})-1}(x_{p}) \;\;.
\eeq
Apart from a condensate 'seed' functional
\(\wt{\Delta}(\hat{T}_{(\Psi);M_{\psi};N_{\psi}}^{\boldsymbol{-1};AB}(x_{p})\,,\,
\hat{T}_{(\Psi);M_{\psi};N_{\psi}}^{AB}(x_{p})\,;\,\hat{J}_{\psi\psi;M_{\psi};N_{\psi}}^{AB}(x_{p})\,)\),
one achieves the square root of the determinant and the bilinear source field part
\(J_{\psi;M_{\psi}}^{T,A}(x_{p})\,\ldots\,J_{\psi;N_{\psi}}^{B}(y_{q})\) with matrix
\(\hat{\mscr{M}}_{M_{\psi};N_{\psi}}^{AB}(x_{p},y_{q})\) (\ref{s4_14}). This matrix
\(\hat{\mscr{M}}_{M_{\psi};N_{\psi}}^{AB}(x_{p},y_{q})\) is composed of the 'Nambu' doubled, kinetic
Hamiltonian part (\ref{s4_15},\ref{s4_17}) with self-energy densities
\(\hat{\mfrak{s}}_{M_{\psi};N_{\psi}}^{(\Psi,\hat{B})(\kappa)B\ppr A\ppr}(x_{p})\),
\(\hat{\mfrak{s}}_{M_{\psi};N_{\psi}}^{(\Psi,\hat{W})(a;\kappa)B\ppr A\ppr}(x_{p})\),
\(\hat{\mfrak{s}}_{M_{q};N_{q}}^{(Q,\hat{G})(\alpha;\kappa)B\ppr A\ppr}(x_{p})\)
from the background averaging functional, the generating source matrix 
\(\hat{\mscr{J}}_{{\scrscr M_{\psi};N_{\psi}}}^{AB}(x_{p};y_{q})\)
for observables and the gradient term 
\(\delta\hat{\mscr{H}}_{\hat{\mathrm{S}}}(\hat{T}_{(\Psi)}^{-1},\hat{T}_{(\Psi)})\) (\ref{s4_16}).
Furthermore, one achieves the effective gauge fields
\(\boldsymbol{\matG}(x_{p})\) (\ref{s4_18}), \(\mscr{G}_{\nu}^{\gamma}(x_{p})\) (\ref{s4_21}),
\(\boldsymbol{\matW}(x_{p})\) (\ref{s4_19}), \(\mscr{W}_{\nu}^{c}(x_{p})\) (\ref{s4_22}),
\(\boldsymbol{\matB}(x_{p})\) (\ref{s4_20}), \(\mscr{B}_{\nu}(x_{p})\) (\ref{s4_23}) consisting of the matrices
\(\hat{\mfrak{M}}_{\beta\mu,\gamma\nu}^{(\hat{G})}\) (\ref{s3_9}),
\(\hat{\mfrak{M}}_{b\mu,c\nu}^{(\hat{W})}\) (\ref{s3_11}), \(\hat{\mfrak{M}}_{\mu,\nu}^{(\hat{B})}\) (\ref{s3_14})
with self-energies for gauge field strength tensors and tilded currents \(\wt{\mfrak{J}}_{\alpha\mu}^{(\hat{G})}(x_{p})\) (\ref{s4_3}),
\(\wt{\mfrak{J}}_{a\mu}^{(\hat{W})}(x_{p})\) (\ref{s4_5}), \(\wt{\mfrak{J}}_{\mu}^{(\hat{B})}(x_{p})\) (\ref{s4_7}) also composed
of self-energy matrices in place of the gauge field strength tensors without any Fermi fields.
A convenient approximation for the background averaging in (\ref{s4_13}) follows from a saddle point computation so that
the effective gauge field variables (\ref{s4_18},\ref{s4_21}), (\ref{s4_19},\ref{s4_22}), (\ref{s4_20},\ref{s4_23})
and self-energy densities 
\(\hat{\mfrak{s}}_{M_{\psi};N_{\psi}}^{(\Psi,\hat{B})(\kappa)B\ppr A\ppr}(x_{p})\),
\(\hat{\mfrak{s}}_{M_{\psi};N_{\psi}}^{(\Psi,\hat{W})(a;\kappa)B\ppr A\ppr}(x_{p})\),
\(\hat{\mfrak{s}}_{M_{q};N_{q}}^{(Q,\hat{G})(\alpha;\kappa)B\ppr A\ppr}(x_{p})\)
are replaced by definite, fixed spacetime
functions \(\langle\boldsymbol{\matG}(x_{p})\rangle\), \(\langle\boldsymbol{\matW}(x_{p})\rangle\),
\(\langle\boldsymbol{\matB}(x_{p})\rangle\) and 
\(\langle\hat{\mfrak{s}}_{M_{\psi};N_{\psi}}^{(\Psi,\hat{B})(\kappa)B\ppr A\ppr}(x_{p})\rangle\),
\(\langle\hat{\mfrak{s}}_{M_{\psi};N_{\psi}}^{(\Psi,\hat{W})(a;\kappa)B\ppr A\ppr}(x_{p})\rangle\),
\(\langle\hat{\mfrak{s}}_{M_{q};N_{q}}^{(Q,\hat{G})(\alpha;\kappa)B\ppr A\ppr}(x_{p})\rangle\).
These fixed spacetime functions are
determined from first order variations of the background averaging functional '\(\langle\ldots\rangle\)' where
the anti-hermitian ingredient of latter background functional has to comply with the already introduced,
various anti-hermitian epsilon terms for a stable, proper convergence of Green functions.

\section{Summary and conclusion} \lb{s5}

\subsection{The combination of the Higgs field with the gauge self-energies} \lb{s51}

Apart from the case of the strong interaction (\ref{s3_9}), the Higgs fields add to various parts,
as to the self-energies of the electroweak interaction in (\ref{s3_11},\ref{s3_14}) and also to their
generalized currents (\ref{s3_12},\ref{s3_15}). Therefore, it may be difficult to observe pure Higgs field
contributions in experiments, in particular because the Higgs field adds to self-energies for the gauge
field parts in various, rather involved combinations. However, it turns out that one conclude for coset matrices
\(\hat{T}_{(\Psi)}(x_{p})=\exp\{-\hat{Y}(x_{p})\}\) in an effective generating functional with effective,
composed gauge fields, determined from a saddle point computation of a background functional. The total
dimension \(N_{0}=90\) for the SSB with \(\mbox{SO}(N_{0},N_{0})\,/\,\mbox{U}(N_{0})\otimes\mbox{U}(N_{0})\)
is the exact, relevant number of the total standard model which is, however, very different from the
large \(\mbox{SU}_{c}(\mbox{N}_{c})\) expansions with the number of colour degrees of freedom from \(\mbox{N}_{c}=3\)
to \(\mbox{N}_{c}\rightarrow\infty\) \cite{Witten1,Witten2,Witten3}. In addition to this aspect
of a large dimension \(N_{0}=90\), it is also of interest to what extent the vanishing axial anomaly
and corresponding vanishing sum of Hopf invariants can constrain the topology of field pairs of fermions or
anomalous self-energy parts within the standard model \cite{Fuji}. 
Hence, it remains to determine and to investigate detailed,
nontrivial field combinations of the coset matrix \(\hat{T}_{(\Psi)}(x_{p})\) with vanishing sum
of Hopf invariants for the anti-symmetric, quaternion eigenvalues and total matrix elements within
the generator \(\hat{Y}(x_{p})\), similar to the Skyrme-Faddeev field theory for the strong interaction case
\cite{Fad1,Manton,BCSQCD}.

In this paper we have demonstrated how to obtain a classical field theory with self-energy matrices in a
coset decomposition for simplifying the total path integral of the quantized standard model of electroweak
interactions. Since one concentrates on the self-energy matrices, representing the irreducible propagator
parts of a quantum many-body theory, our derived classical field theory also has a semiclassical notion.
The given approach of this article and of Ref. \cite{BCSQCD} generalizes to super-symmetric QCD and to
the super-symmetric case of electroweak forces which will be considered separately. The various steps of
sections \ref{s3} and \ref{s4} can be directly transferred to super-symmetric cases where one can perform
a similar anomalous doubling of Fermi fields, but with inclusion of similar 'Nambu' doubling for the
bosonic partners in the total chiral super-fields.

\end{document}